\documentclass[epsfig,psfig,aps,twocolumn,prl,showpacs]{revtex4-1}
\usepackage{hyperref} 
\hypersetup{colorlinks,allcolors=red}
\usepackage{amsfonts,amsmath,bbm,wasysym}
\usepackage{amsfonts,dsfont}
\usepackage[english]{babel}
\usepackage{graphicx}
\usepackage{epsfig}
\usepackage{relsize}
 \newcommand{\sgn}{\operatorname{sgn}} 
\begin{document}

\title{Stacking-induced Chern insulator}

\author{Marwa Manna\"i$^1$}
%\email{}
\author{Jean-No\"el Fuchs$^2$}
\email{fuchs@lptmc.jussieu.fr}
\author{Frédéric Piéchon$^3$}
\email{frederic.piechon@universite-paris-saclay.fr}
\author{Sonia Haddad$^{1,4}$}
\email{sonia.haddad@fst.utm.tn}
\affiliation{
$^1$Laboratoire de Physique de la Mati\`ere Condens\'ee, Facult\'e des Sciences de Tunis, Universit\'e Tunis El Manar, Campus Universitaire 1060 Tunis, Tunisia\\
$^2$Sorbonne Universit\'e, CNRS, Laboratoire de Physique Th\'eorique de la Mati\`ere Condens\'ee, LPTMC, 75005 Paris, France\\
$^3$Université Paris-Saclay, CNRS, Laboratoire de Physique des Solides, 91405, Orsay, France\\
$^4$ Institute for Solid State Physics, University of Tokyo, Kashiwa, Chiba 277-8581, Japan
}

\date{\today}

\begin{abstract}
Graphene can be turned into a semimetal with broken time-reversal symmetry by adding a valley-dependent pseudo-scalar potential that shifts the Dirac point energies in opposite directions, as in the modified Haldane model. We consider a bilayer obtained by stacking two time-reversed copies of the modified Haldane model, where conduction and valence bands cross to give rise to a nodal line in each valleys. In the AB stacking, the interlayer hopping lifts the degeneracy of the nodal lines and induces a band repulsion, leading surprisingly to a chiral insulator with a Chern number $C=\pm2$. As a consequence, a pair of chiral edge states appears at the boundaries of a ribbon bilayer geometry. In contrast, the AA stacking does not show nontrivial topological phases. We discuss possible experimental implementations of our results.

\end{abstract}

\maketitle

\section{Introduction} A Chern insulator~\cite{Hasan-Rev,Qi-Rev,Bansil} is a two-dimensional (2D) topological insulator with broken time reversal symmetry (TRS), where chiral edge states emerge in a ribbon geometry with counterpropagating directions at the opposite boundaries of the strip. These edge states are the hallmark of the bulk topological properties described by a topological invariant, the Chern number $C$, indicating the number of the chiral edge channels.\
A Chern insulator exhibits quantum anomalous Hall effect (QAHE)~\cite{AQHE-rev,Yang,Khan,Ni,Esslinger,Kee,acoustic,Serlin1,Serlin2,resistance,tuneC} introduced by Haldane in his seminal paper~\cite{Haldane}.\

The Haldane model~\cite{Haldane} (HM) describes a honeycomb lattice with complex hopping integrals between next nearest-neighbors (NNN), creating staggered magnetic fluxes, which break TRS. The NNN hopping terms are characterized by a complex phase $\Phi$ which has the same sign in the two sublattices of the honeycomb lattice. \

\begin{figure}[hpbt] 
\begin{center}
\includegraphics[width=1\columnwidth]{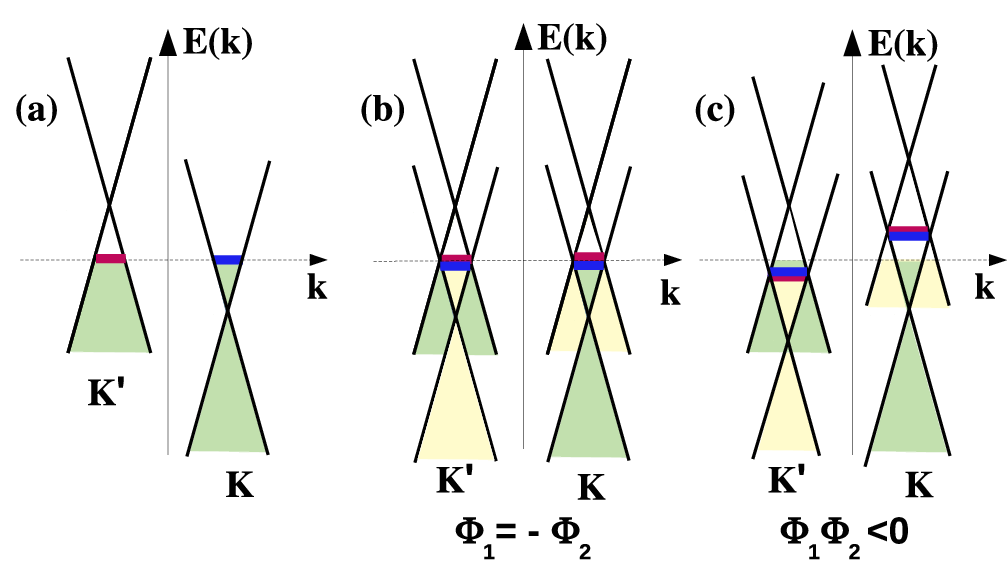}
\end{center}
\caption{(a) Fermi surface of the monolayer mHM at half filling. A valley-dependent scalar potential shifts, oppositely, the energies of the Dirac cones. The Fermi surface consists into an electron pocket (blue line) in one valley and a hole pocket (red line) in the opposite valley. (b) and (c) Nodal lines (red and blue lines) of the mHM on uncoupled bilayers where the intralayer pseudo-scalar potential have opposite signs corresponding (b) to $\Phi_1=-\Phi_2$, and (c) to phases of opposite signs and unequal magnitudes ($\Phi_1\Phi_2<0$), where $\Phi_l$, ($l=1,2$) is the intralayer complex phase of the NNN hopping integrals. }
\label{Fig-intro}
\end{figure}

A modified Haldane model (mHM) has been proposed in Ref.~\onlinecite{Franz} where TRS is broken by a valley-dependent pseudo-scalar potential which shifts, oppositely, the energies of the Dirac points in the two valleys. The potential is generated by the sign flip of the complex phase $\Phi$ in one of the honeycomb sublattices. The system turns into a semi-metal with a Fermi surface consisting, at half filling, of a hole pocket, in a valley, and an equal-sized electron pocket in the opposite valley (Fig.~\ref{Fig-intro}(a)). 
The so-called antichiral edge states \cite{Franz} are expected to emerge in zigzag nanoribbons described by the mHM: they are unidirectional gapless edge modes that co-propagate at the opposite ribbon boundaries and are counterbalanced by bulk states.
\newline\

Imagine, now, a bilayer structure of the HM where the layers, indexed by $l=1,2$, host Haldane phases labeled by the Chern numbers $C_l$.  A naive intuition tells us that the total Chern number is $C=C_1+C_2$, as reported in Ref.~\onlinecite{Dutta} for the AA bilayer. In particular, a vanishing Chern number $C=0$ is expected if the layers have opposite chiralities resulting from oppositely broken TRS~\cite{Dutta} (see Appendix \ref{App-HM}).
This expectation is at the heart of the Kane and Mele~\cite{Kane} idea, where the layer index is replaced by the spin projection and the two opposite TRS copies of the spin polarized HM give rise to a vanishing ($\mathbb{Z}$) Chern number. The latter leaves room to another ($\mathbb{Z}_2$) TRS protected topological invariant corresponding to a quantized spin Hall insulator. 
 
In this paper, we consider two stacked layers of the mHM, where the semi-metallic layers break TRS in opposite ways (Fig.~\ref{Fig-intro}) and have undefined Chern number $C_l=\emptyset$ ($l=1,2$). The resulting bilayer structure shows counter-intuitive behaviors. Depending on the precise stacking order (AA, AB or BA), we find that the resulting system may be gapless (AA) with $C=\emptyset$ or topologically gapped with a Chern number $C\neq C_1+C_2$ but $C= \pm 2$ (AB/BA). In other words, and in contrast with the HM bilayer, the stacking order in the mHM bilayer is a key parameter controlling the gap opening and the emergence of chirality. Understanding the origin of the topological gapped phases and the corresponding chirality is the main objective of the present work.
%An experimental implementation of our results can be achieved in photonic crystals~\cite{Private-pho}, electrical circuit lattices~\cite{Private-elec} and {\color{blue} bilayer of hexagonal transition metal dichalcogenides (TMD)~\cite{Franz}} . \

The paper is organized as follow. In section II, we derive the Bloch Hamiltonian of a generic AB (BA) bilayer structure where the layers are time reversed copies of the mHM. We show that the system turns, under a finite interlayer hopping, to an insulator belonging to the symmetry class A \cite{AZ}. In section III, we derive, based on a perturbative approach in the large interlayer coupling limit, the analytical expression of the Chern number $C$ of this insulating phase which is found to be $C=\pm 2$, in agreement with our numerical calculations. To highlight the presence of the chiral edge states of the $C=\pm2$ phase, we present, in section IV, numerical electronic band structures calculated in a ribbon geometry of the mHM bilayers. In Section V, a heuristic argument is presented to explain 
the emergence of chirality in the mHM AB (BA) bilayer, and its absence in the AA bilayer. In Section VI, we discuss possible experimental realizations of our findings in real and artificial materials. The concluding section VII summarizes our results. The paper also contains five appendices providing detailed numerical and analytical results.

%%%%%
\section{Bilayer modified Haldane model} 
\begin{figure}[hpbt] 
\begin{center}
\includegraphics[width=1\columnwidth]{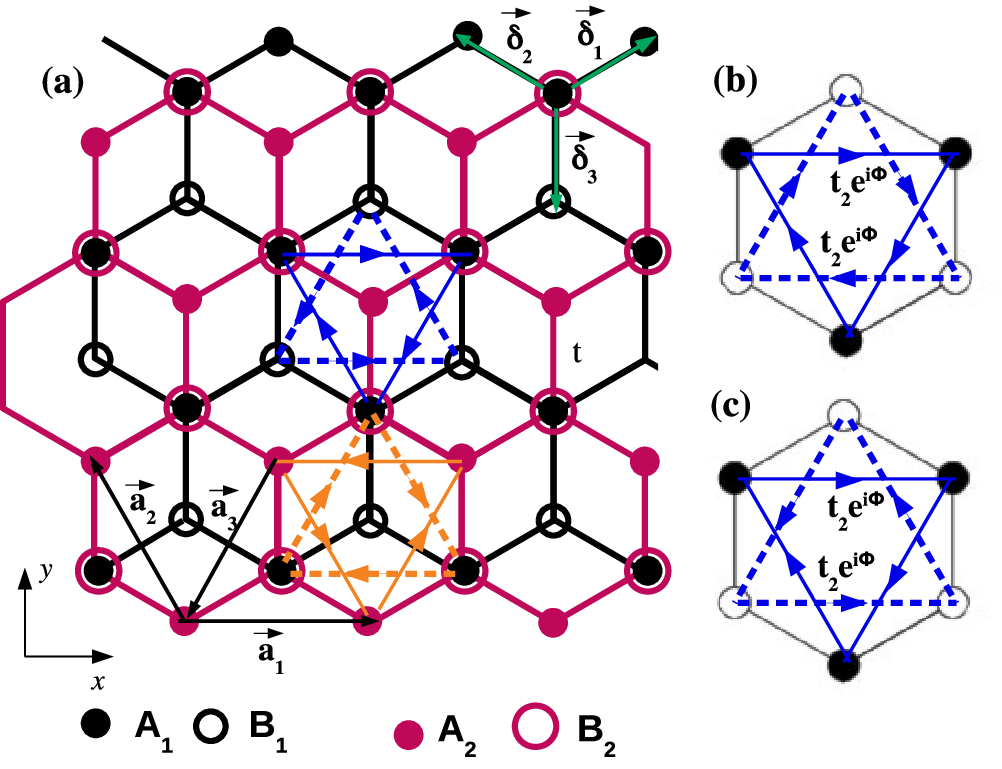}
\end{center}
\caption{(a) Modified Haldane model on AB stacked bilayer. The monolayer unit cell atoms are denoted by $A_l$ and $B_l$, where $l=1,2$ is the layer index.
 $\boldsymbol{\delta}_i$ ($i=1,2,3$) are the vectors connecting nearest neighboring (NN) atoms,
$\left(\mathbf{a}_1,\mathbf{a}_2\right)$ is the lattice basis and $\Phi$ is the phase of the complex NNN hoppings. The solid (dashed) arrows indicate the pattern of the complex NNN hopping terms in the sublattice $A$ ($B$). The blue and orange lines correspond, respectively, to layer $1$ and $2$.
(b) NNN hopping in the Haldane model and in (c) the modified Haldane model. }
\label{mHM-lattice}
\end{figure}

We start with the HM on a honeycomb lattice with a unit cell containing two different types of atoms denoted A and B. The corresponding spinless fermionic Hamiltonian is~\cite{Haldane}
\begin{eqnarray}
 H_{\text{H}}=t\sum_{\langle i,j\rangle } c_i^{\dagger}c_j+t_2\sum_{\langle\langle i,j \rangle\rangle} e^{-i\Phi_{ij}}c_i^{\dagger}c_j
 +\sum_i M_i c_i^{\dagger}c_i,
\end{eqnarray}
where $t$ ($t_2$) is the hopping integral to first (second) nearest neighbors, $c_i$ annihilate a spinless fermion on atom ($i$), $\Phi_{ij}=\nu_{ij}\Phi$ is the complex phase of NNN hopping integrals and $\nu_{ij}=\pm 1$ according to the pattern given in Fig.~\ref{mHM-lattice} (b).
The last term describes the Semenoff masses where $M_i=M$ ($-M$) for A (B) atoms.\

%\begin{widetext}
%\end{widetext}
In the sublattice pseudo-spin basis, the HM Bloch Hamiltonian can be written as
\begin{eqnarray}
H_{\text{H}}(\mathbf{k})=a^0_{\mathbf{k}}\sigma_0+b_{\mathbf{k}}\sigma_x+c_{\mathbf{k}}\sigma_y+\left(a_{\mathbf{k}}+M\right)\sigma_z ,
\label{HMmL}
\end{eqnarray}
where $\boldsymbol{\sigma}$ and $\sigma_0$ are, respectively, the sublattice Pauli and $2\times2$ identity matrices, 
$ b_{\mathbf{k}}=\Re\left(f_{\mathbf{k}}\right)$, $ c_{\mathbf{k}}=-\Im\left(f_{\mathbf{k}}\right)$, $f_{\mathbf{k}}=t\sum^3_{i=1} e^{i\mathbf{k}\cdot\boldsymbol{\delta}_i}$, $a^0_{\mathbf{k}}=2t_2\cos\Phi\sum^3_{i=1} \cos\left(\mathbf{k}\cdot\mathbf{a}_i\right)$ and $a_{\mathbf{k}}=-2t_2\sin\Phi\sum^3_{i=1} \sin\left(\mathbf{k}\cdot\mathbf{a}_i\right)$ is the Haldane mass.
 The vectors
$\boldsymbol{\delta}_i$  ($i=1,2,3$) connect an atom to its first neighbors and ($\mathbf{a}_1,\mathbf{a}_2$) is the Bravais lattice basis given by (Fig.~\ref{mHM-lattice} (a)): $\mathbf{a}_1=\sqrt{3}a\mathbf{e}_x,\,
\mathbf{a}_2=-\frac{\sqrt{3}}2a\mathbf{e}_x+a\frac 32\mathbf{e}_y$, where $a$ is the distance between nearest neighbors. We also define $\mathbf{a}_3=-\left(\mathbf{a}_1+\mathbf{a}_2\right)$ (see Appendix \ref{App-HM}). 

The modified Haldane model can be derived from the HM (Eq.~\ref{HMmL}) by flipping the sign of the phase $\Phi$ on one sublattice, as shown in Fig.~\ref{mHM-lattice}~(c). The corresponding Hamiltonian~\cite{mHM symm1} can be deduced from Eq.~\ref{HMmL} by changing $a_\mathbf{k} \sigma_z$ by $a_\mathbf{k}\sigma_0$
\begin{eqnarray}
H_{\text{mH}}(\mathbf{k})=\left(a_{\mathbf{k}}+a^0_{\mathbf{k}}\right)\sigma_0+b_{\mathbf{k}}\sigma_x+c_{\mathbf{k}}\sigma_y+M\sigma_z,
\label{mHMmL}
\end{eqnarray}
which describes a semi-metal (due to band overlap) for $|M|\leq M_c\equiv 3\sqrt{3}t_2 \sin\Phi$, and a trivial insulator otherwise.
In the gapless phase, antichiral edge states emerge at the boundaries of a ribbon structure~\cite{Franz}.\

Since $a_{\mathbf{k}}=-a_{-\mathbf{k}}$ for $\Phi\not\equiv 0\mod{[\pi]}$, the HM and the mHM break TRS as $\mathcal{T}^{\dagger} H_{\alpha}(\mathbf{k})\mathcal{T}\neq H_{\alpha}(-\mathbf{k})$, where $\alpha=\text{HM, \,mHM}$, $\mathcal{T}=K$ and $K$ denotes complex conjugation.
Charge conjugation $\mathcal{C}=\sigma_zK$ and the sublattice chiral symmetries $\mathcal{S}=\sigma_z$ are also broken as 
$\mathcal{C}^{\dagger} H_{\alpha}(\mathbf{k})\mathcal{C}\neq -H_{\alpha}(-\mathbf{k})$ and 
$\mathcal{S}^{\dagger} H_{\alpha}(\mathbf{k})\mathcal{S}\neq -H_{\alpha}(\mathbf{k})$.
The HM belongs to the A class of topological insulators ~\cite{AZ} characterized by a $\mathbb{Z}$ invariant (Chern number)~\cite{HM symm} while the mHM is a semi-metal or a trivial insulator as stated before.\\

As the AA-stacked mHM bilayer does not show nontrivial topological behavior (see Appendix \ref{App-mHM-AA}), we consider the AB-stacked bilayer and denote by $\Phi_l$ ($l=1,2$) the complex phase in layer ($l$) (see Appendix \ref{App-AB-BA}).\
The mHM in Bernal bilayer, is described by the following Bloch Hamiltonian written, in the basis of the four orbitals forming the unit cell ($A_1,B_1,A_2,B_2$) (Fig.~\ref{mHM-lattice}~(a)) as
\begin{widetext}
\begin{eqnarray}
 H_{\text{B}}(\mathbf{k})=
\begin{pmatrix}
A_{1,\mathbf{k}}+M_1 & f_{\mathbf{k}} &0& 2t_{\perp}\\
f^{\ast}_{\mathbf{k}} & A_{1,\mathbf{k}}-M_1 &0&0\\
0&0&A_{2,\mathbf{k}}+M_2 & f_{\mathbf{k}}\\
2t_{\perp}&0&f^{\ast}_{\mathbf{k}} & A_{2,\mathbf{k}}-M_2 
\end{pmatrix} ,
\label{HBL}
\end{eqnarray}
\end{widetext}
where we only considered the interlayer coupling $2t_{\perp}$ between dimer sites $\left(A_1,B_2\right)$. Here $M_l$ ($l=1,2$) is the layer Semenoff mass, and $A_{l,\mathbf{k}}=a_{l,\mathbf{k}}+a^0_{l,\mathbf{k}}$, where 
\begin{eqnarray}
a_{l,\mathbf{k}}&=&-2t_2\sin\Phi_l\sum_{i=1}^3 \sin\left(\mathbf{k}\cdot\mathbf{a}_i\right),\nonumber\\
a^0_{l,\mathbf{k}}&=&2t_2\cos\Phi_l\sum_{i=1}^3  \cos\left(\mathbf{k}\cdot\mathbf{a}_i\right).
\label{al}
\end{eqnarray}
Introducing the layer pseudospin $\boldsymbol{\tau}$ Pauli matrices and the corresponding identity matrix $\tau_0$, $H_{B}(\mathbf{k})$ (Eq.~\ref{HBL}) reduces to
\begin{eqnarray}
H_{B}(\mathbf{k})&=&\left( b_{\mathbf{k}}\sigma_x+c_{\mathbf{k}}\sigma_y\right)\tau_0
+2t_{\perp}\left(\sigma_+\tau_+ +\sigma_-\tau_-\right)\nonumber\\
&+&\frac 12 \left(A_1+A_2\right)\sigma_0\tau_0+\frac 12 \left(A_1-A_2\right)\sigma_0\tau_z\nonumber\\
&+&\frac 12 \left(M_1+M_2\right)\sigma_z\tau_0+\frac 12 \left(M_1-M_2\right)\sigma_z\tau_z,
\label{HBL2}
\end{eqnarray} 
where $\sigma_{\pm}=\frac12\left(\sigma_x\pm i\sigma_y\right)$ and $\tau_{\pm}=\frac12\left(\tau_x\pm i\tau_y\right)$.\

This Hamiltonian has some similarity with the Hamiltonian given by Eq. 1 in Ref.~\onlinecite{SOC}, where the authors have studied a modified Kane and Mele~\cite{Kane} model of a graphene layer on a substrate.
The Hamiltonian of Ref.~\onlinecite{SOC} belongs to the class AII~\cite{AZ,AZ2} characterized by a $\mathbb{Z}_2$ invariant, whereas the Hamiltonian given by Eq.~\ref{HBL2} belongs to class A~\cite{AZ,AZ2} labeled by a $\mathbb{Z}$ invariant.\
Indeed, since $a_{l,-\mathbf{k}}=-a_{l,\mathbf{k}}$ (Eq.~\ref{al}), the system breaks TRS, $\mathcal{T}=K$, the charge conjugation represented by $\mathcal{C}=\sigma_z\tau_0K$, with $\mathcal{C}^2=\mathds{1}$ and the chirality $\mathcal{S}=\tau_0\sigma_z$~\cite{footnote sym}.\

To discuss the topological class of the system, one needs to analyze the presence of gaps in the energy spectrum of the Hamiltonian $H_{B}(\mathbf{k})$ (Eq.~\ref{HBL2}) and, in particular, for vanishing Semenoff masses ($M_l=0,\, l=1,2$) where both layers are semi-metals.\

We consider, for simplicity, the case where $\Phi_1=-\Phi_2=\pm\frac{\pi}2$ to drop the energy shift term $a^0_{l,\mathbf{k}}$ (Eq.~\ref{al}) which does not affect the band topology.
In this case, the energy spectrum of $H_{B}(\mathbf{k})$ (Eq.~\ref{HBL2}) shows a particle-hole symmetry~\cite{footnote Eg} and is given by
\begin{eqnarray}
 E_{\alpha_1,\alpha_2}(\mathbf{k})=\alpha_1\sqrt{A_{\mathbf{k}}+\alpha_2 B_{\mathbf{k}}},
 \label{E+-}
\end{eqnarray}
where $\alpha_i=\pm 1$ and
\begin{eqnarray}
 A_{\mathbf{k}}=a^2_{\mathbf{k}}+|f_{\mathbf{k}}|^2+2t_{\perp}^2,
 B_{\mathbf{k}}=2\sqrt{|f_{\mathbf{k}}|^2\left(a^2_{\mathbf{k}}+t_{\perp}^2\right)+t_{\perp}^4} . \nonumber\\
 \label{ABk}
\end{eqnarray}
The eigenenergies given by Eq.~\ref{E+-} obey the inequalities
$E_{-,+}(\mathbf{k})\leq E_{-,-}(\mathbf{k})\leq E_{+,-}(\mathbf{k})\leq E_{+,+}(\mathbf{k})$ and the gap separating the energy bands around the zero-energy is $\Delta=\min_{\mathbf{k}}\left(\Delta_{\mathbf{k}}\right)=2\sqrt{A_{\mathbf{k}}-B_{\mathbf{k}}}$. 
The gap closing condition is
\begin{eqnarray}
\left(|f_{\mathbf{k}}|^2-a^2_{\mathbf{k}}\right)^2=-4a^2_{\mathbf{k}} t_{\perp}^2,
\label{gap}
\end{eqnarray}
which can be satisfied in two cases: (i) for $a_{\mathbf{k}}=0$ ($t_2=0$), which corresponds to an AB graphene bilayer with only nearest neighboring (NN) hopping terms and a quadratic contact point between the bands at zero energy~\cite{McCann}; 
(ii) for uncoupled layers ($t_{\perp}=0$) where Eq.~\ref{gap} reduces to $|a_{\mathbf{k}}|=|f_{\mathbf{k}}|$, which defines two non-intersecting closed loops in the Brillouin zone (BZ) and results into two valley closed Fermi lines, originating from two overlapping bands (Fig.~\ref{Fig-intro}). 
The gap (Eq.~\ref{gap}) is therefore finite as soon as $t_\perp$ and $t_2$ are non-zero~\cite{footnote}.\

At this point, a question arises: what are the possible values of the Chern number $C$ of the gapped phases resulting from the instability of the Fermi lines? Numerically, we find that $C=\pm 2$ when the two phases $\Phi_l\in\left[-\pi,\pi\right] $ ($l=1,2$) are of opposite signs ($\Phi_1\Phi_2<0$) and that the system is gapless when they are of the same sign.\

In order to understand these findings, we proceed in two steps. We first derive an effective $2\times 2$ Hamiltonian in the limit of large interlayer coupling ($t_2\ll t_{\perp}$) at which it is simple to get an analytical expression of the Chern number of the highest filled energy band, denoted $E_{-,-}(\mathbf{k})$ (Eq.~\ref{E+-}) around half-filling. 
Restricting the analytical calculations to this limit is fully justified in the case of vanishing Semenoff masses since the topology of the system is unchanged when the ratio $t_2/t_{\perp} $ crosses from $t_2/ t_{\perp}\ll 1$ to the opposite limit $t_2/ t_{\perp}\gg 1$.
The reason is that the gap does not close as soon as $a_{\mathbf{k}}$ or/and $t_\perp$ are turned on (Eq.~\ref{gap}), which prevents any topological phase transition.\

In a second step, we calculate the energy-spectrum of the mHM on AB-stacked nanoribbons to bring out the signature of the chiral edge states corresponding to the bulk Chern insulating phase.\\

\section{Effective two-band model }
In the limit of large interlayer coupling ($t_2\ll t_{\perp}$), the energy bands corresponding to the dimer $(A_1,B_2)$, coupled by $t_{\perp}$, are pushed away from zero-energy. The lowest energy bands around half-filling can then be described by an effective $2\times 2$ model written in the subspace of the uncoupled orbitals $A_2$ and $B_1$.\

To derive the low energy Hamiltonian, we use the L\"owdin partitioning method~\cite{Lowdin,McCann}. The effective $2\times 2$ Hamiltonian reduces to (see Appendix \ref{App-mHM-model})
\begin{widetext}
\begin{eqnarray}
H_\text{eff}({\mathbf{k}})=
-2t_{\perp}\frac{f_{\mathbf{k}}^2}{X^2}\sigma_+
-2t_{\perp}\frac{f_{\mathbf{k}}^{\ast2}}{X^2}\sigma_- 
+\left[\frac 12\left(M_1+M_2\right)
-a_{\mathbf{k}}\left(1-\frac{|f_{\mathbf{k}}|^2}{X^2}\right)\right]\sigma_z
+\frac 12\left(M_2-M_1\right)\sigma_0
\label{Heff}
\end{eqnarray}
\end{widetext}
where the $\boldsymbol{\sigma}$ Pauli matrices are now written in the $(A_2, B_1)$ basis and $X^2=(a_\mathbf{k}+M_1)(a_\mathbf{k}+M_2)+4 t_\perp^2$.\

For $M_1=M_2=0$, the eigenenergies are $E_{\text{eff},\pm}(\mathbf{k})=\pm\sqrt{ a^2_{\mathbf{k}}\left(1-\frac{|f_{\mathbf{k}}|^2}{X^2}\right)^2+4t^2_{\perp}\frac{|f_{\mathbf{k}}|^4}{X^4}}$. $E_{\text{eff},-}$ is equal, to the leading order in $\frac{|f_{\mathbf{k}}|}{t_{\perp}}$, to $E_{--}(\mathbf{k})$ (Eqs.~\ref{E+-}, \ref{ABk}).\

In order to characterize the topology of the Hamiltonian given by Eq.~(\ref{Heff}), we consider the limit $M_l,\, t_2\ll t_{\perp}$ ($l=1,2$) and expand $H_\text{eff}$ around the Dirac points $\xi \mathbf{K}$, where $\xi=\pm$ is the valley index, so that Eq.~\ref{Heff} becomes (see Appendix \ref{App-mHM-model})
\begin{widetext}
\begin{eqnarray}
H_\text{eff}(\mathbf{q})=-\frac{\hbar^2 v^2}{t_{\perp}}\left(q^2_x-q^2_y\right)\sigma_x
+2\tau_z\frac{\hbar^2v^2}{t_{\perp}}q_xq_y\sigma_y+ \left[\frac 12\left(M_1+M_2\right)-\sgn\left({\Phi_1}\right) 3\sqrt{3}t_2\tau_z\right]\sigma_z+\frac 12\left(M_2-M_1\right)\sigma_0 ,
\label{Heff3}
\end{eqnarray} 
\end{widetext}
where $\mathbf{q}=\mathbf{k}-\xi \mathbf{K}$ and $\Phi_1=-\Phi_2=\pm \frac{\pi}{ 2}$. The two first terms in Eq.~\ref{Heff3} describe the low energy Hamiltonian of a Bernal bilayer graphene~\cite{McCann} while the $\sigma_z$ term contains both a Haldane $a_{\mathbf{k}}\sim 3\sqrt{3}t_2\xi$ and a Semenoff $\frac 12\left(M_1+M_2\right)$ mass terms.\

The Chern number associated to the lowest band of this two-band Hamiltonian is~\cite{Sticlet}
\begin{eqnarray}
C=\sum_{\xi} \frac 12 \chi \sgn(m_{\xi}),
\label{Chern-def}
\end{eqnarray}
where $m_{\xi}=\frac 12\left(M_1+M_2\right)-\sgn\left({\Phi_1}\right) 3\sqrt{3}t_2\xi$ is the total mass and $\chi=-2\xi$ is the chirality of the quadratic band contact point~\cite{McCann}.\

For vanishing Semenoff masses ($M_l=0$) where topological phase transitions are prohibited, $\sgn(m_{\xi})=-\sgn\left({\Phi_1}\right) \xi$, which gives $C=\sgn\left({\Phi_1}\right) 2$.
This result is in agreement with our numerical calculations of the Chern number $C$ of the four energy bands $E_{\alpha_1,\alpha_2}(\mathbf{k})$ (Eqs.~\ref{E+-}, \ref{ABk}): $C_{-,+}=0$, $C_{-,-}=\pm 2$, $C_{+,-}=\mp2$, and $C_{+,+}=0$ giving rise, for the occupied bands, to a total Chern number $C=2 \,(-2)$ for $\Phi_1=-\Phi_2=\frac{\pi}2(-\frac{\pi}2)$.

The chiral insulating phases occur in the case where the complex phase $\Phi_1$ and $\Phi_2$ have opposite signs and the Chern number of the lowest energy band is defined as far as the system is gapped (see Appendix \ref{App-mHM-phi-mass}.

\section{Zigzag and armchair ribbons} 
\begin{widetext}
\begin{figure}[hpbt] 
\begin{center}
$\begin{array}{cc}
\includegraphics[width=0.5\columnwidth]{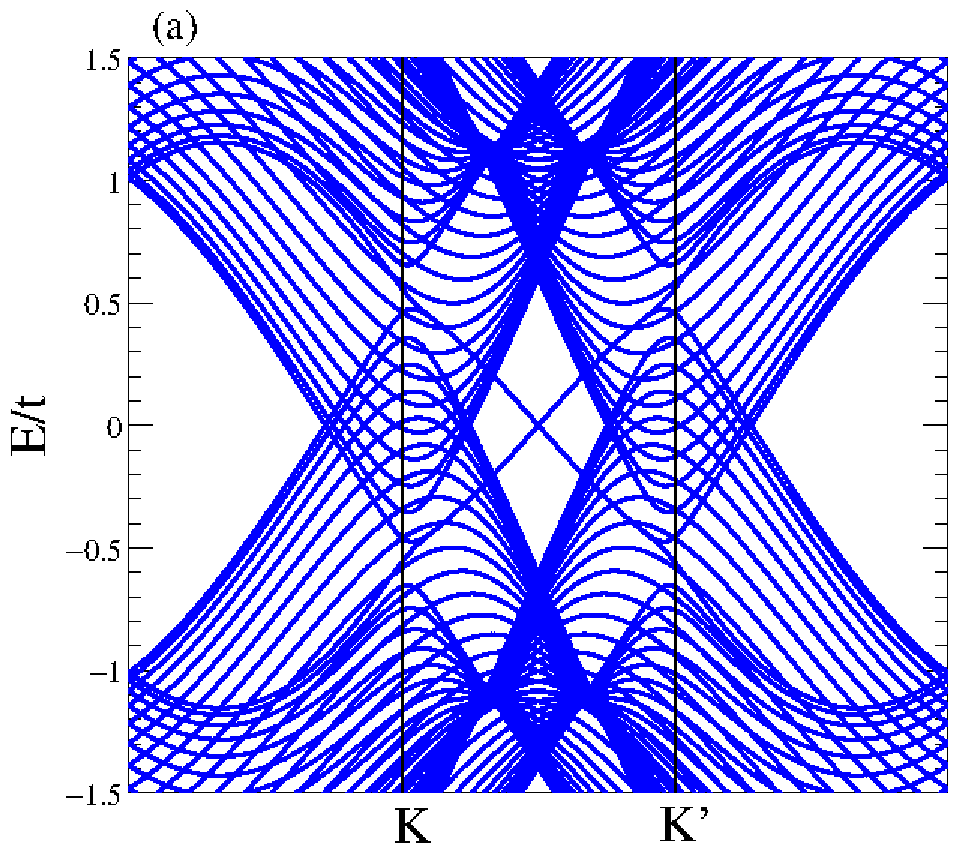}&
\includegraphics[width=0.5\columnwidth]{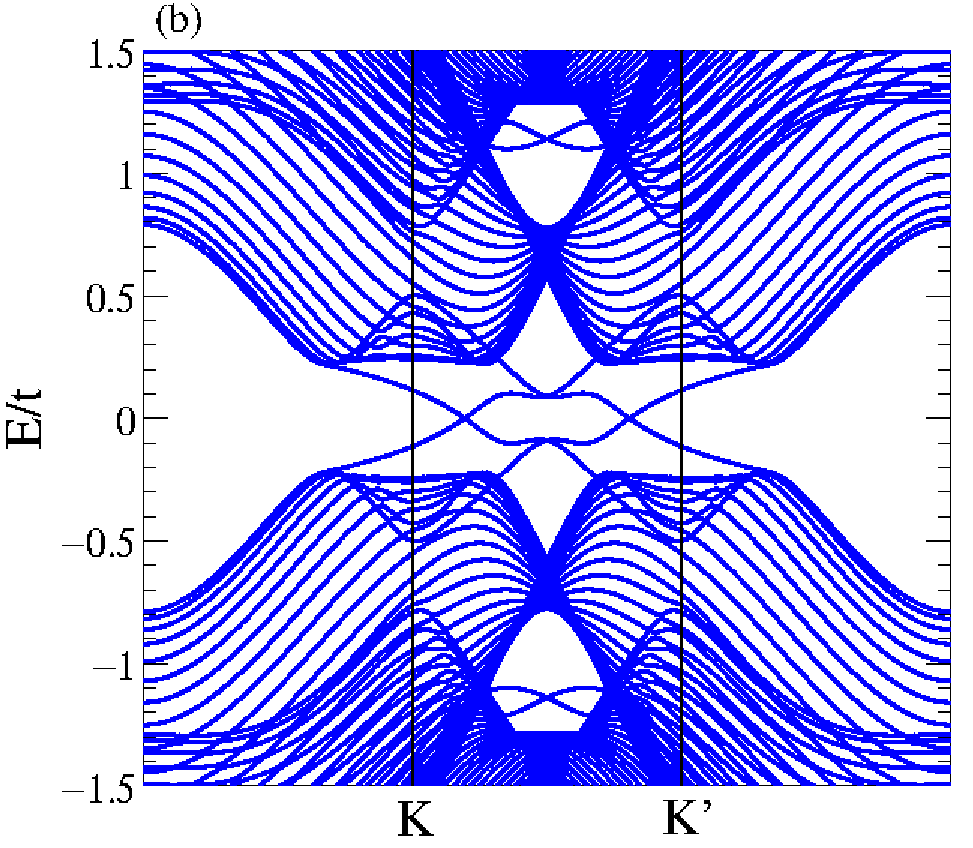}\\
\includegraphics[width=0.55\columnwidth]{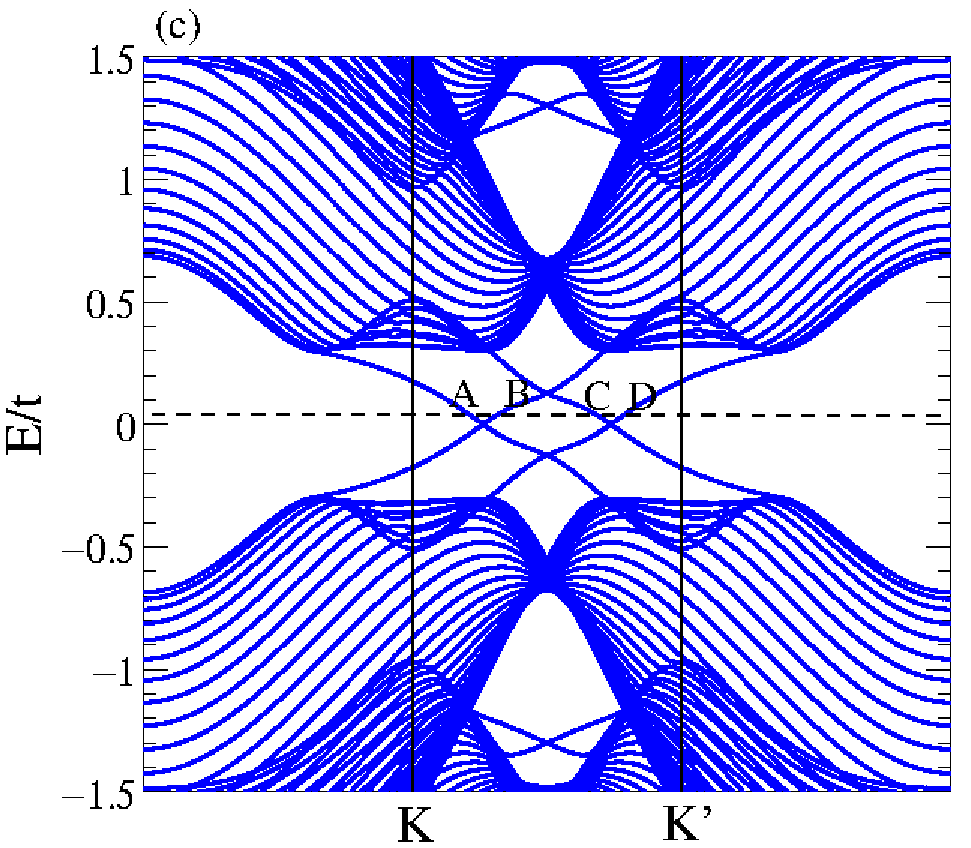}&
\includegraphics[width=0.45\columnwidth]{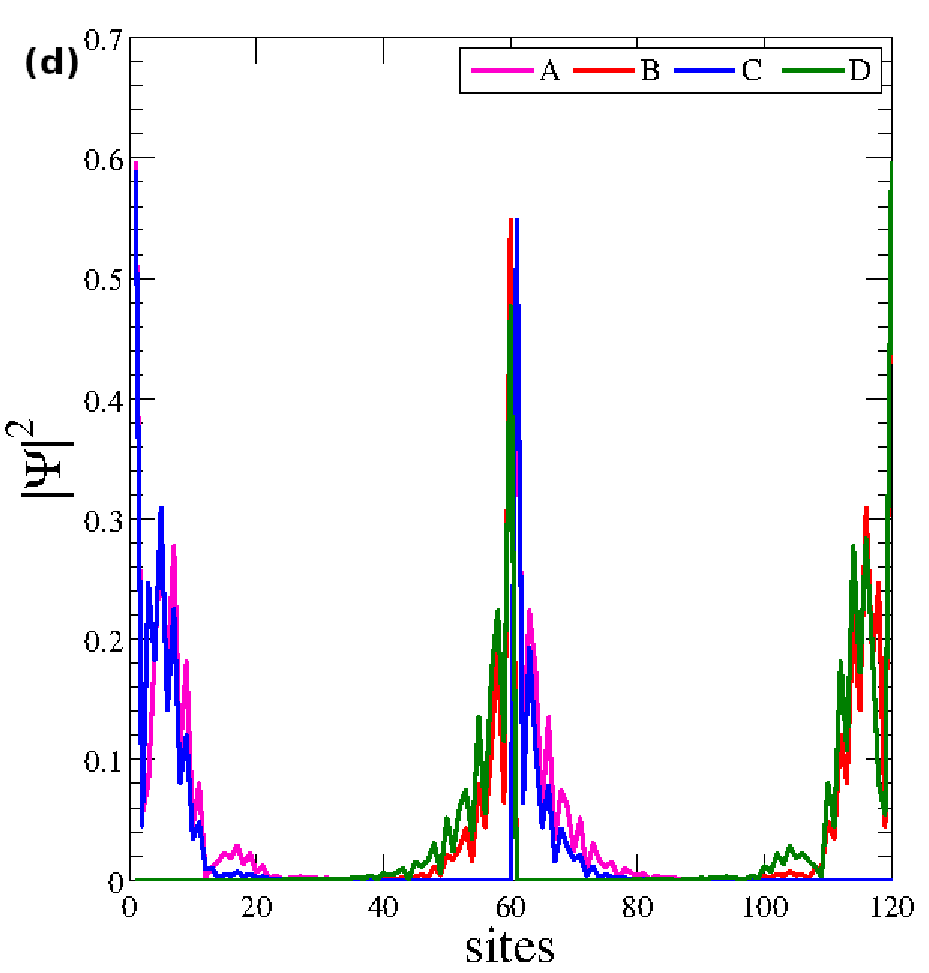}
\end{array}$
\end{center}
\caption{Band structure of an AB bilayer mHM on zigzag nanoribbons of a width $W = 60$ atoms. Calculations are done for $t_2 = 0.1t$, $\Phi_1=-\Phi_2= \frac{\pi}2$, and $M_1=M_2=0$. The interlayer hopping is (a) $t_{\perp}=0$, (b) $t_{\perp}=0.5t$ and (c) $t_{\perp}=0.8t$. (d) Probability distributions of the edge states denoted by A, B, C and D in (c) located at the energy indicated by the dashed line. Sites numbered 1 to 60 (resp. 61 to 120) belong to the first (resp. second) layer.}
\label{bandZZ}
\end{figure}
\end{widetext}

Figure~\ref{bandZZ} shows the band structure of AB stacked nanoribbons with zigzag boundaries described by the mHM with complex phases $\Phi_1=-\Phi_2=\frac{\pi}2$. 
For a non vanishing interlayer coupling $t_{\perp}$, a gap opens at half filling and two pairs of chiral edge states emerge. \

The corresponding probabilities are depicted in Fig.~\ref{bandZZ} (d) indicating that the two edge modes, appearing at the left side of the ribbon, have roughly equal weight on the two layers and have the same velocity. Similarly, the right boundary also supports two edge modes, but counterpropagating with the left side channels.
This feature confirms that the coupled semi-metal ribbons turn into a topological Chern insulator with a Chern number $C=2$. By flipping the signs of the complex phases $\Phi_1=-\Phi_2=-\frac{\pi}2$, the  direction of propagation of the chiral edge states at each boundary is reversed, which results into a Chern number $C=-2$. \

\begin{figure}[hpbt] 
\begin{center}
$\begin{array}{cc}
\includegraphics[width=0.5\columnwidth]{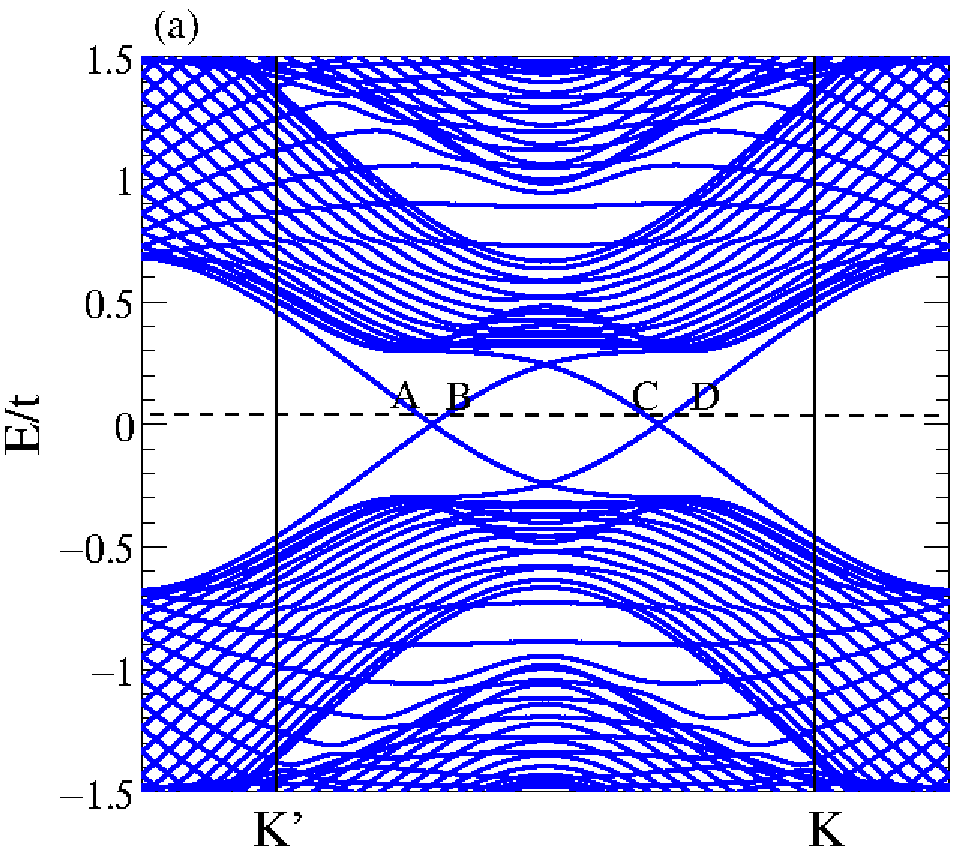}
\includegraphics[width=0.5\columnwidth]{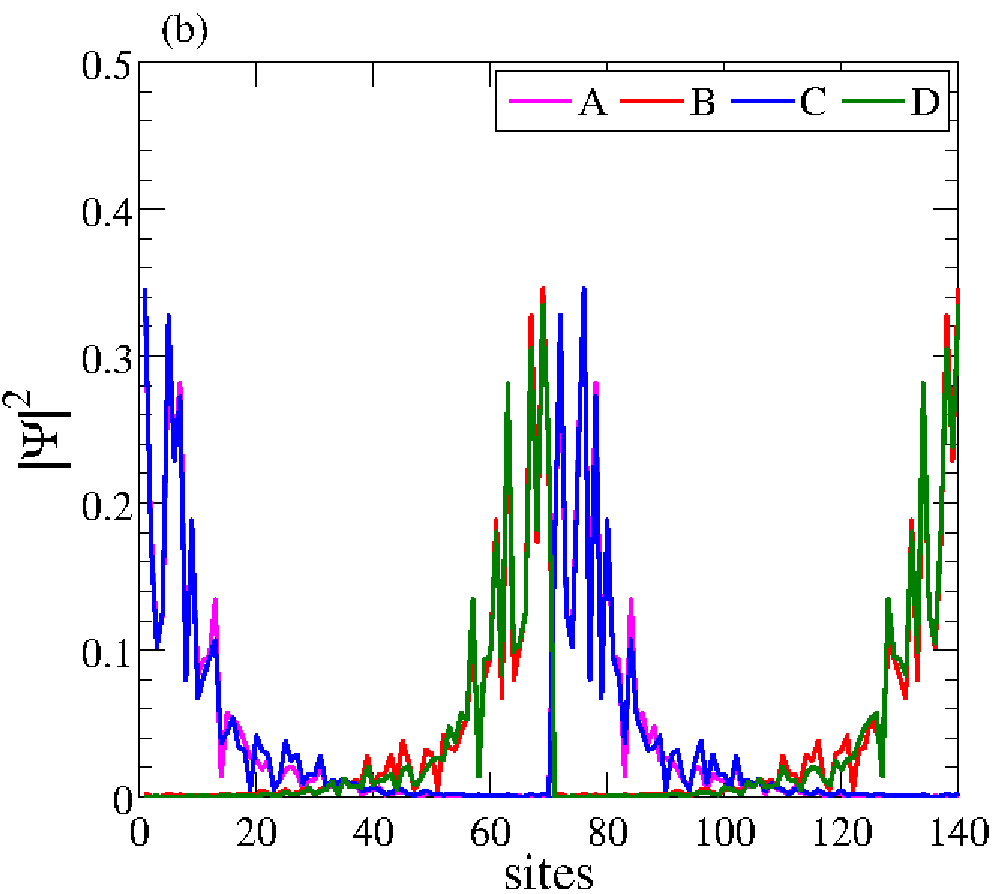}
\end{array}$
\end{center}
\caption{(a) Band structure of an AB bilayer mHM on armchair nanoribbons of a width $W = 70$ atoms and for $t_2 = 0.1t$, $t_{\perp}=0.8t$, $M_1=M_2=0$ and $\Phi_1=-\Phi_2= \frac{\pi}2$. (b) Corresponding probability distributions of the edge states denoted by A, B, C and D in (a). Sites numbered 1 to 70 (resp. 71 to 140) belong to the first (resp. second) layer.}
\label{AC}
\end{figure}

In figure~\ref{AC}, we represent the band structure of the mHM bilayer ribbons with armchair boundaries in the case of $\Phi_1=-\Phi_2=\frac{\pi}2$ and for vanishing Semenoff masses. Under the interlayer coupling, the system becomes a $C=2$ Chern insulator.\

The analytical expression of the Chern number (Eq.~\ref{Chern-def}) is derived in the limit $M_l,\, t_2 \ll t_{\perp}$. To go beyond this limit, we perform numerical calculations for different values of $M_l$, $\Phi_l$, $t_2$ and $t_{\perp}$ (see Appendix \ref{App-mHM-phi-mass}).\

For uncoupled layers, the closed Fermi lines survive if the layers remain semi-metallic, which is the case for $|M_l|\leq M_{lc}\equiv 3\sqrt{3}t_2| \sin\Phi_l| $, ($l=1,2$), where the critical mass $M_{lc}$ marks the transition between the semi-metallic phase ($M_l\le M_{lc}$) to the gapped phase ($M_l>M_{lc}$) of the monolayer mHM. By turning on the interlayer coupling, our numerical calculations show that, for realistic hopping integrals ($t_{\perp}\sim t_2$) the system is gapped and becomes a Chern insulator. The corresponding Chern number is $C=\pm 2$ regarding the presence of two channels of chiral edge states at each boundary of a bilayer strip.
If the Semenoff mass overcomes the Haldane mass $|M_{lc}|<|M_l|$, the monolayers are trivial insulators and the interlayer hopping $t_{\perp}$ brings the mHM bilayer to a trivial gapped phase with a vanishing Chern number $C=0$, regardless of the stacking order (AA, AB or BA) (see Appendix \ref{App-mHM-phi-mass}).\\

In the case of the AA bilayer mHM, the system turns into a semi-metal (trivial insulator) in the absence (presence) of Semenoff mass terms, regardless of the nature of the ribbon's boundaries (see Appendix \ref{App-mHM-AA}).\

\section{Stacking-induced chirality}

\begin{figure}[hpbt] 
\begin{center}
$\begin{array}{c}
\includegraphics[width=0.8\columnwidth]{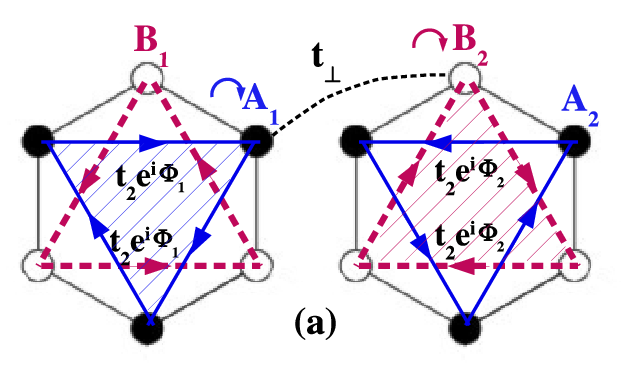}\\
\includegraphics[width=0.8\columnwidth]{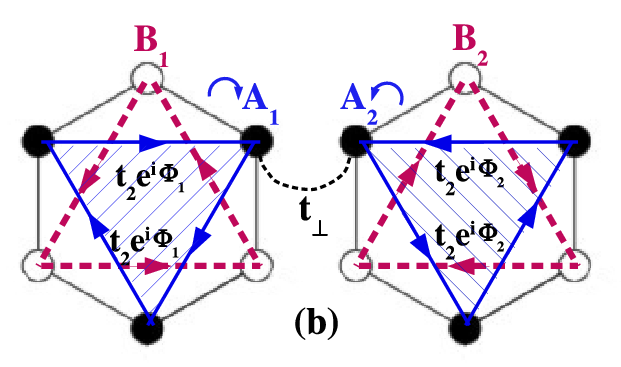}
\end{array}$
\end{center}
\caption{Schematic interpretation of the stacking-dependent chirality of the mHM bilayers in the case of opposite complex phases $\Phi_1=-\Phi_2$. (a) In the AB stacking, the interplane hopping $t_{\perp}$ couples the fluxes of $A_1$ and $B_2$ sublattices flowing in the same direction, giving rise to a dominant clockwise chirality and a pair of chiral edge states in each layer. (b) In the AA stacked layers, the $A_1$ ($B_1$) and the $A_2$ ($B_2$) sublattices have opposite fluxes with no dominant chirality.}
\label{AB-AA}
\end{figure}

We now explain how the chirality emerges due to the stacking order of the layers. This feature can be understood from the schematic representation of the interlayer hopping illustrated in Fig.~\ref{AB-AA}, where we consider the case of opposite complex phases $\Phi_1=-\Phi_2$. Imagine a situation of finite $t_2$ and of slowly turning on $t_\perp$ in order to see the emergence of a Chern (resp. trivial) insulator in the case of AB (resp. AA) stacking. \
In AB-stacked bilayers, the interlayer hopping couples the sublattice $A_1$ and $B_2$ where the fluxes flow in the same direction (see Fig.~\ref{AB-AA}(a)). Consequently, the dimer chirality dominates and gives rise to a pair of chiral edge states at each layer end. \
If the interlayer hopping concerns the $B_1$ and the $A_2$ atoms (BA-stacking), the flux flows are flipped in comparison with the case where the dimer is $\left(A_1,B_2\right)$ and the system gains an opposite chirality.\

However, in the AA stacked bilayer mHM, the $A_1$ and the $A_2$ sublattices have opposite fluxes which cannot result in a dominant chirality (see Fig.~\ref{AB-AA} (b)), and the system, if gapped, cannot support chiral edge states  (see Appendices \ref{App-HM} and \ref{App-mHM-AA}).\\

When stacking a layer with its time-reversed copy, the intuition is that time-reversal symmetry should be restored and therefore the bilayer should not be chiral. This is indeed what occurs for AA stacking  (see Appendix \ref{App-mHM-AA}). However, AB/BA stacking favors one chirality by explicitly breaking the symmetry between the two layers.\

One could also imagine a situation in which AA stacking becomes unstable and spontaneously chooses between AB and BA stacking. A possible mechanism would be similar to a Peierls instability, in which a loss in elastic (lattice deformation) energy is compensated by a gap opening leading to a gain in electronic energy. The net result would be a spontaneous time-reversal symmetry breaking and the spontaneous emergence of a Chern insulator (see e.g. Ref.~\onlinecite{SC} for a similar idea). We leave such a study to future work.\\

\section{Possible experimental realizations}
How to implement experimentally the stacking-induced $C=\pm 2$ Chern insulator in bilayer of mHM ?
The experimental realization of such a phase depends on the state-of-the-art of the implementation of the monolayer mHM in real systems.\

As shown in Fig.6, the realization of this intriguing Chern insulator requires complex phases with opposite signs ($\Phi_1\Phi_2<0$) and not necessary $\Phi_1=-\Phi_2=\pi/2$  which we have considered in our calculations for the sake of simplicity (Fig. \ref{phi}).\

A first implementation of such phase could be achieved in electric circuits \cite{Private-elec,Yang21} and photonic crystals \cite{Private-pho,Zhou} where two layers of artificial mHM, with complex phases of opposite signs, are coupled by an interlayer tunneling.\

It is worth noting that the monolayer mHM has not yet been realized in real materials.
Colomés and Franz \cite{Franz} proposed the hexagonal transition metal dichalcogenides (TMD) monolayers, and in particular WSe$_2$, as excellent candidates to realize the mHM.\

Based on this idea, we propose that a Bernal bilayer of WSe$_2$ may be a platform to realize the $C=\pm2$ stacking induced Chern insulator.\

To have opposite signs for the complex phases in both layers ($\Phi_1\Phi_2<0$), we propose that one layer should be hole-doped while the other be electron-doped. This requirement could be understood from Fig.4(a) of Ref.\onlinecite{Franz} showing the band structure of WSe$_2$ nanoribbon where the edge states in the valence band (VB) and the conduction band (CB) have opposite group velocities. If the Fermi level crosses the edge state of the VB (CB), the system may mimic a monolayer mHM with a positive (negative) complex phase $\Phi$ (Fig. \ref{phi}) (see Appendix \ref{App-mHM-phi-mass}).\

We then expect to realize the Chern $C=\pm2$ gapped phase in an AB stacked bilayer of a hole doped (h-WSe$_2$) and an electron-doped (e-WSe$_2$).
The electron and hole doping of WSe$_2$ have already been achieved using substitutional doping \cite{Mukherjee} and field induced electron doping \cite{Chen}.\\

What are the experimental fingerprints of the $C=\pm2$ Chern phase in bilayer stacked mHM?\

The Hall resistance is expected to be quantized as $R_{xy}=\pm 2 \frac h{e^2}$ \cite{Serlin1}. Dissipationless transport properties of the chiral edge states could also be used to probe the emergence of the chiral modes \cite{Ying}.\

Scanning tunneling microscopy (STM) has been widely used to map the gapless edge states of topological materials \cite{Zahid}. Within this technique, the differential tunneling conductance $\frac{dI}{dV}$, which measures the local density of states, is expected to show a pronounced step-edge within the gap of the h-WSe$_2$/e-WSe$_2$ AB-bilayer, indicating the presence of chiral edge states crossing the Chern gap \cite{Zahid}.
Moreover, STM in finite magnetic field \cite{Choi} can be, also, used to uncover the topological nature of the gapped phase of h-WSe$_2$/e-WSe$_2$ AB-bilayer and the corresponding Chern number based on the Landau fan diagram.
Local compressibility measurements using scanning single electron transistor \cite{Pablo2} are a powerful probe to detect the incompressible chiral edge states and to encode their Chern number indexation.
Atomic force microscopy \cite{Kim} and angle-resolved photo-emission spectroscopy \cite{Claudia} are also possible techniques to reveal the presence of the chiral edge states.

\section{Conclusion} 

\begin{figure}[hpbt] 
\begin{center}
\includegraphics[width=1\columnwidth]{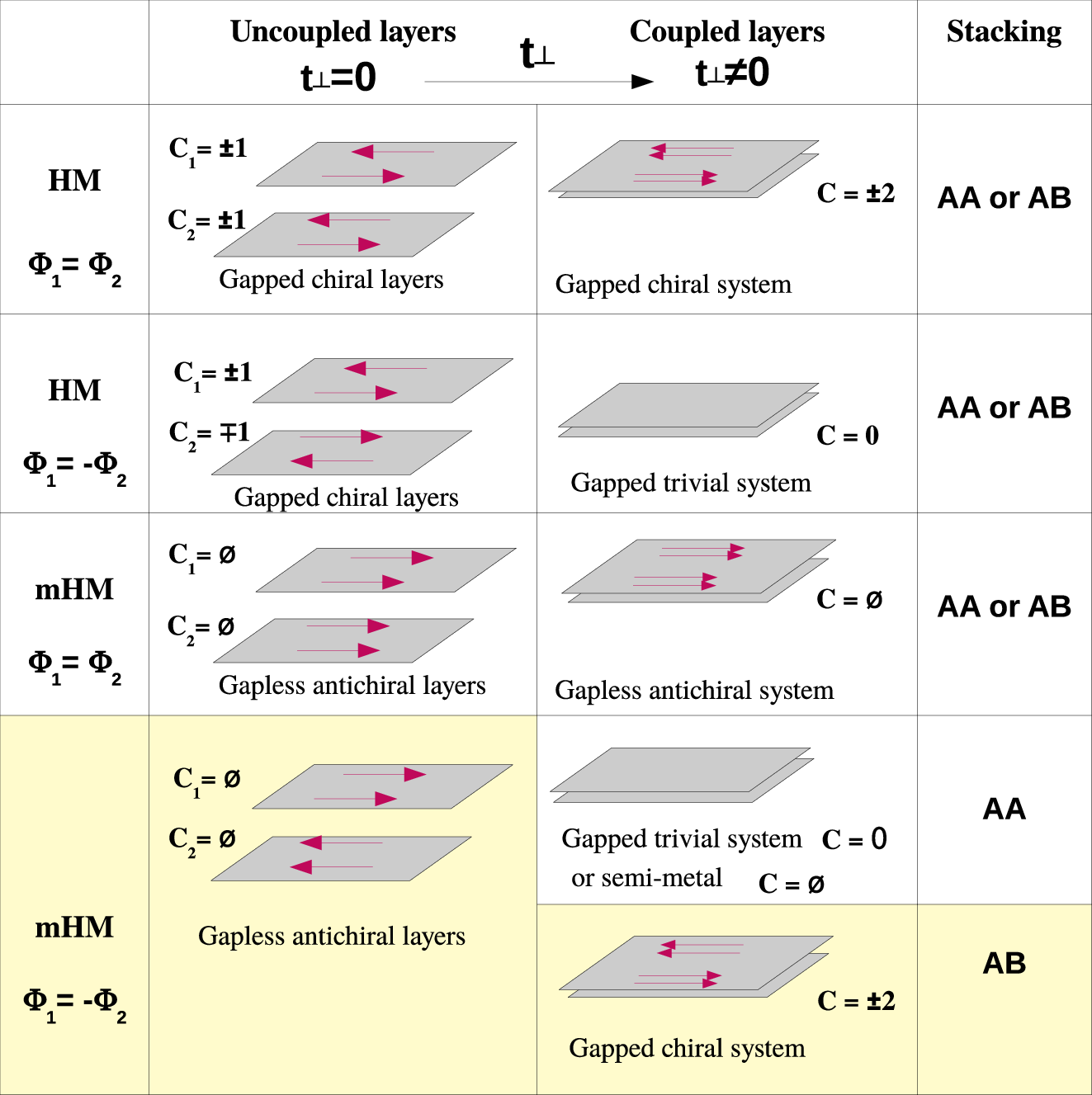}
\end{center}
\caption{Summary of the stacking-dependent properties of the Haldane (HM) and the modified Haldane (mHM) models in bilayer honeycomb lattices.
$\Phi_l$ ($l=1,2$) is the complex phase of the NNN hopping term in layer $l$, 
$C_l$ is the corresponding Chern number while $C$ is the total Chern number of the lowest occupied bands at half-filling. $C=\emptyset$ means an undefined Chern number. The yellow cells indicate the case showing unexpected topological behavior (i.e. $C\neq C_1+C_2$), which has been studied in the main text. The other cases are discussed in the appendices \ref{App-HM}, \ref{App-mHM-AA} and \ref{App-mHM-phi-mass}.
\label{table-res}}
\end{figure}

We discussed the topological properties of the modified Haldane model~\cite{Franz} in a bilayer of honeycomb lattice where the TRS is broken, oppositely, in the uncoupled semi-metallic layers. We found that, in the AB stacked bilayer, the interlayer hopping drives the system into a topological insulator, with a Chern number $C=\pm2$. The smoking gun of this insulating phase is the systematic emergence of two channels of chiral edge states at the boundaries of the AB bilayer ribbons, regardless of the boundary nature (zigzag/armchair) of the ribbons~\cite{bearded,bearded-Ref}. However, the modified Haldane model in the AA stacked bilayer is found to be a semi-metal or a trivial insulator depending on the value of the Semenoff masses. Our results are schematically summarized in Fig.~\ref{table-res}, where we also give the behavior of the HM in AA and AB stacked bilayers.\
This stacking-induced Chern insulator could be readily realized in bilayers of electrical circuits~\cite{Private-elec} and photonic crystals~\cite{Private-pho} by coupling two time reversal copies of the mHM which was already implemented in a microwave-scale gyromagnetic photonic crystal~\cite{Zhou} and in electrical circuits~\cite{Yang21}.\
An experimental implementation with real material can be achieved in the hexagonal $\text{WSe}_2$, where the Dirac cone shift is due to spin orbit coupling~\cite{Franz}. We propose that a Chern insulator with $C=\pm 2$ can be hosted by a Bernal bilayer of $\text{WSe}_2$ where one layer is hole-doped while the other is electron-doped.
The present work could be extended to multilayer structures of semi-metals with broken TRS, opening the way to tunable Chern insulators, as realized with heterostructures of topological insulators ~\cite{Ctunue}.
We also expect a twist-induced Chern insulator~\cite{Guy} in a moiré superlattice of twisted honeycomb bilayers~\cite{Herrero}, where AA stacked domains form a triangular lattice alternating with AB and BA regions. A gapless (or a gapped trivial insulating) state, emerging in the AA domains, may coexist with Chern insulating phases in the AB and the BA regions with, respectively, a Chern number $C=\pm2$ and $C=\mp2$~\cite{Guy}.

%%%%%%%%%%%%%%%%%%%%%%%%%%%%%%%%%%%%%%%%%%%%%%%%%%%%%%%%%%%%%%%%%%%%%%%%%%%%%%%%%%%%%%%%
\acknowledgements
This work was supported by the Tunisian Ministry of Higher Education and Scientific Research. S. H. acknowledges the LPS in Orsay for financial support and kind hospitality. J.N.F. acknowledges financial support from Institut de Science des Matériaux (iMAT) at Sorbonne Université. We thank  A. Meszaros for fruitful discussions and a critical reading of the manuscript. We also acknowledge B. Zhang, Y. Chong and Y. Yang for discussions about the experimental implementation of our results.

%%%%%%%%%%%%%%%%%%

%Supplemental Material

%\widetext
%\section*{\large Supplemental material for stacking-induced Chern insulator}
\appendix
\section{A. AA and AB bilayer Haldane model}
\label{App-HM}
\renewcommand{\thefigure}{A\arabic{figure}}
\setcounter{figure}{0}

We consider a HM bilayer with AB or AA stackings where the layers are assumed to have complex NNN phases $\Phi_1=-\Phi_2=\frac{\pi}2$, to drop the global energy shift $a^0_{\mathbf{k}}=0$ (Eq.~\ref{al} of the main text) which does not affect the topology of the system.
In the basis of the four orbitals forming the unit cell ($A_1,B_1,A_2,B_2$) the corresponding Hamiltonians can be written as
\begin{eqnarray}
 H_\text{AA-HM}(\mathbf{k})=
\begin{pmatrix}
a_{\mathbf{k}} +M_1 & f_{\mathbf{k}} &2t_{\perp}&0 \\
f^{\ast}_{\mathbf{k}} & -a_{\mathbf{k}} -M_1&0&2t_{\perp}\\
2t_{\perp}&0&-a_{\mathbf{k}} +M_2 & f_{\mathbf{k}}\\
0&2t_{\perp}&f^{\ast}_{\mathbf{k}} & a_{\mathbf{k} -M2}
\end{pmatrix},\nonumber\\
\end{eqnarray}

\begin{eqnarray}
 H_\text{AB-HM}(\mathbf{k})=
\begin{pmatrix}
a_{\mathbf{k}} +M_1 & f_{\mathbf{k}} &0 &2t_{\perp} \\
f^{\ast}_{\mathbf{k}} & -a_{\mathbf{k}} -M_1&0&0\\
0&0&-a_{\mathbf{k}} +M_2 & f_{\mathbf{k}}\\
2t_{\perp}&0&f^{\ast}_{\mathbf{k}} & a_{\mathbf{k} -M2}
\end{pmatrix}.\nonumber\\
\end{eqnarray}
These Hamiltonians can be expressed, using the sublattice and the layer pseudospin matrices $\boldsymbol{\sigma}$ and $\boldsymbol{\tau}$, as
\begin{eqnarray}
&&H_\text{AA-HM}(\mathbf{k})=\left( b_{\mathbf{k}}\sigma_x+c_{\mathbf{k}}\sigma_y\right)\tau_0
+2t_{\perp}\sigma_0\tau_x+ a_{\mathbf{k}}\sigma_z\tau_z \nonumber\\
&&+\frac 12 \left(M_1+M_2\right)\sigma_z\tau_0+\frac 12 \left(M_1-M_2\right)\sigma_z\tau_z,\\
&&H_\text{AB-HM}(\mathbf{k})=\left( b_{\mathbf{k}}\sigma_x+c_{\mathbf{k}}\sigma_y\right)\tau_0
+t_{\perp}\left(\sigma_x\tau_x-\sigma_y\tau_y\right)+ a_{\mathbf{k}}\sigma_z\tau_z\nonumber\\
&&+\frac 12 \left(M_1+M_2\right)\sigma_z\tau_0+\frac 12 \left(M_1-M_2\right)\sigma_z\tau_z.
\label{AB-HM-mass}
\end{eqnarray}
where $a_{\mathbf{k}}$ is given by Eq.~\ref{al} in the main text.\

$H_\text{AA-HM}$ ($H_\text{AB-HM}$) breaks TRS $\mathcal{T}=K$, the charge conjugation, represented by $\mathcal{C}=\sigma_z\tau_zK$ ($\mathcal{C}=\sigma_z\tau_0K$) with $\mathcal{C}^2=\mathds{1}$, and the chirality $\mathcal{S}=\tau_z\sigma_z$ ($\mathcal{S}=\tau_0\sigma_z$).\

In the following, we will show, based on numerical band structure calculations on bilayer ribbons, that coupling two HM with opposite chiralities ($C_1=-C_2$), resulting from oppositely broken TRS ($\Phi_1=-\Phi_2$), gives rise, as expected, to a trivial Chern insulator with $C=C_1+C_2=0$.\
We will discuss the stacking order, the nature of the ribbon edges (zigzag or armchair) and the effect of the intralayer Semenoff masses $M_l$, where $l=1,2$ is the layer index. The case of AA stacking was discussed in Ref.~\onlinecite{Dutta} for a fixed value of the mass term $M_l$.\

\begin{figure}[hpbt] 
\begin{center}
$
\begin{array}{ccc}
\includegraphics[width=0.33\columnwidth]{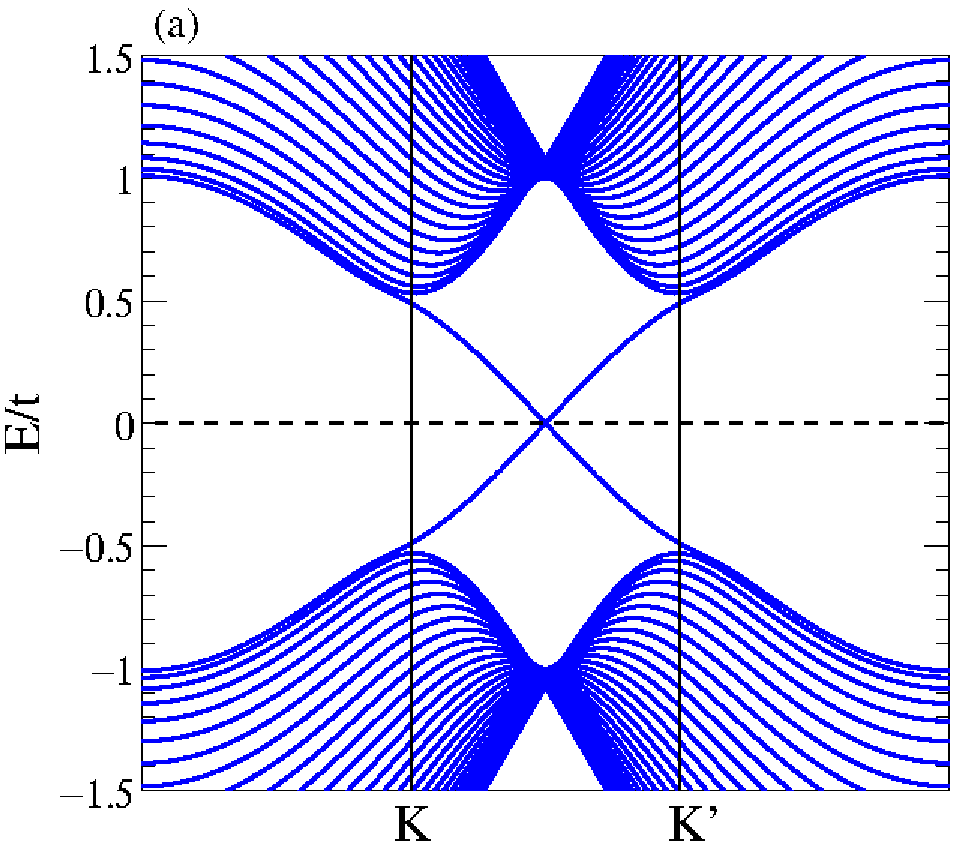}
\includegraphics[width=0.33\columnwidth]{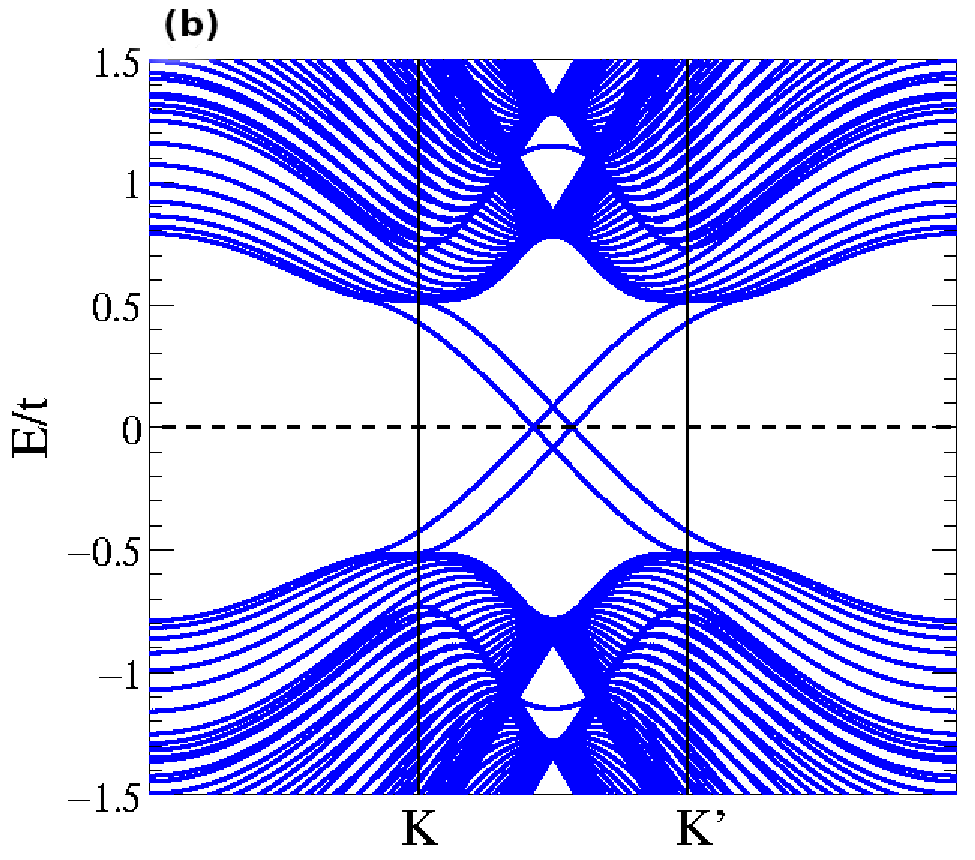}
\includegraphics[width=0.33\columnwidth]{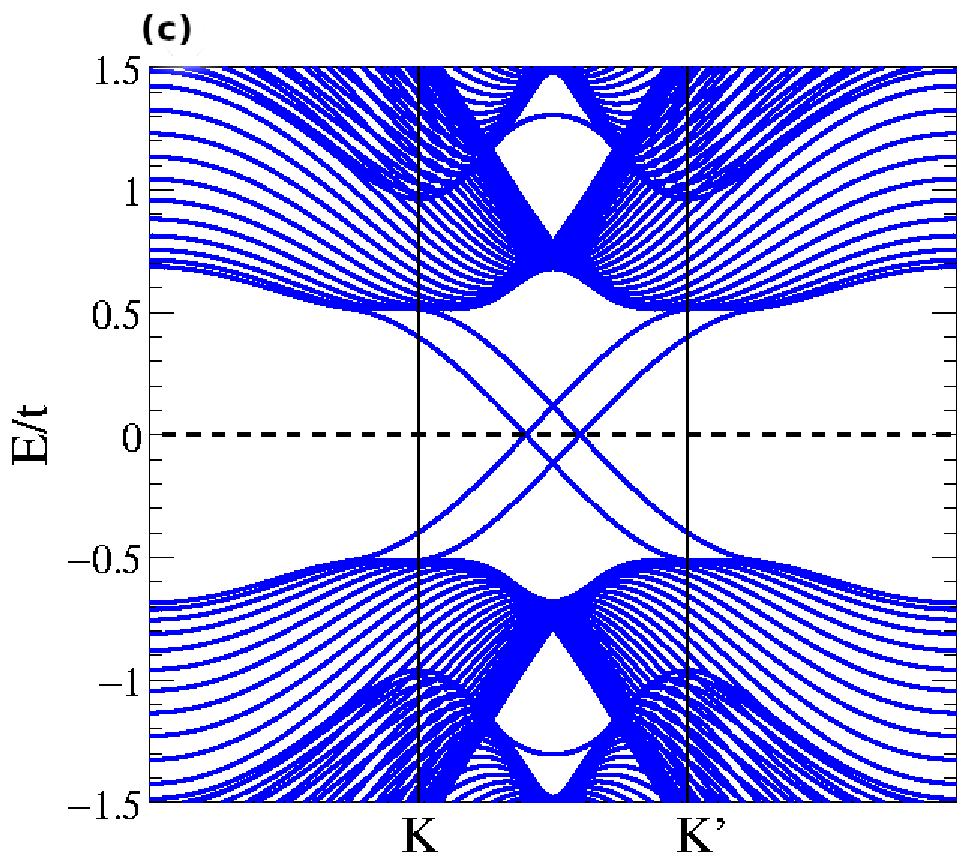}
\end{array}
$
\end{center}
\caption{Tight binding calculations of the electronic band structure of an AB bilayer HM on zigzag nanoribbons of a width $W = 60$ atoms. The interlayer hopping is (a) $t_{\perp}=0$, 
(b) $t_{\perp}=0.5t$ and (c) $t_{\perp}=0.8t$.
Calculations are done for $\Phi_1=\Phi_2= \frac{\pi}2$, $M_1=M_2=0$ and $t_2 = 0.1t$, where $t$ is the NN hopping integral.}
\label{band-HM-ZZ}
\end{figure}

Figure~\ref{band-HM-ZZ} shows the band structure of the AB bilayer HM on zigzag ribbons for $\Phi_1=\Phi_2= \frac{\pi}2$, $M_1=M_2=0$ and at different values of the interlayer hopping $t_{\perp}$. Starting from uncoupled ($t_{\perp}=0$) chiral layers, with equal Chern number $C_1=C_2=\pm 1$, the system turns, under the interlayer coupling, into a Chern insulator with a Chern number $C=\pm 2$ characterized by a pair of chiral edge states propagating at the boundaries of each layer as shown in Fig.~\ref{table-res} of the main text.\
 
\begin{figure}[hpbt] 
$
\begin{array}{ccc}
\includegraphics[width=0.33\columnwidth]{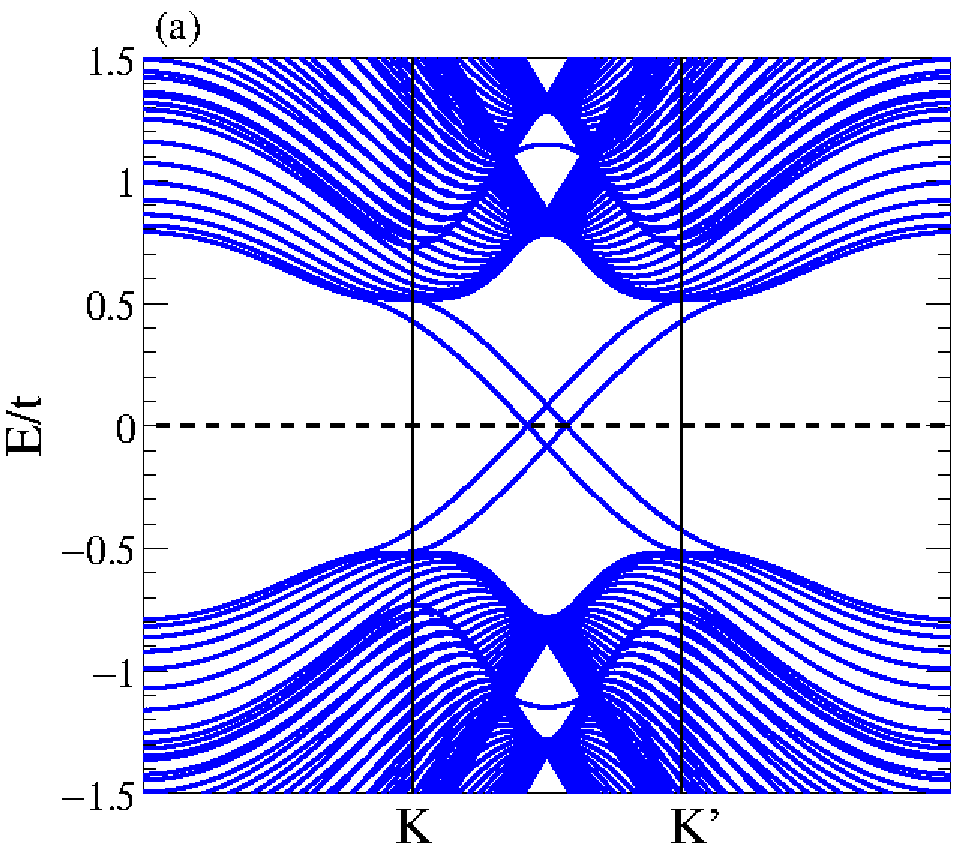}
\includegraphics[width=0.33\columnwidth]{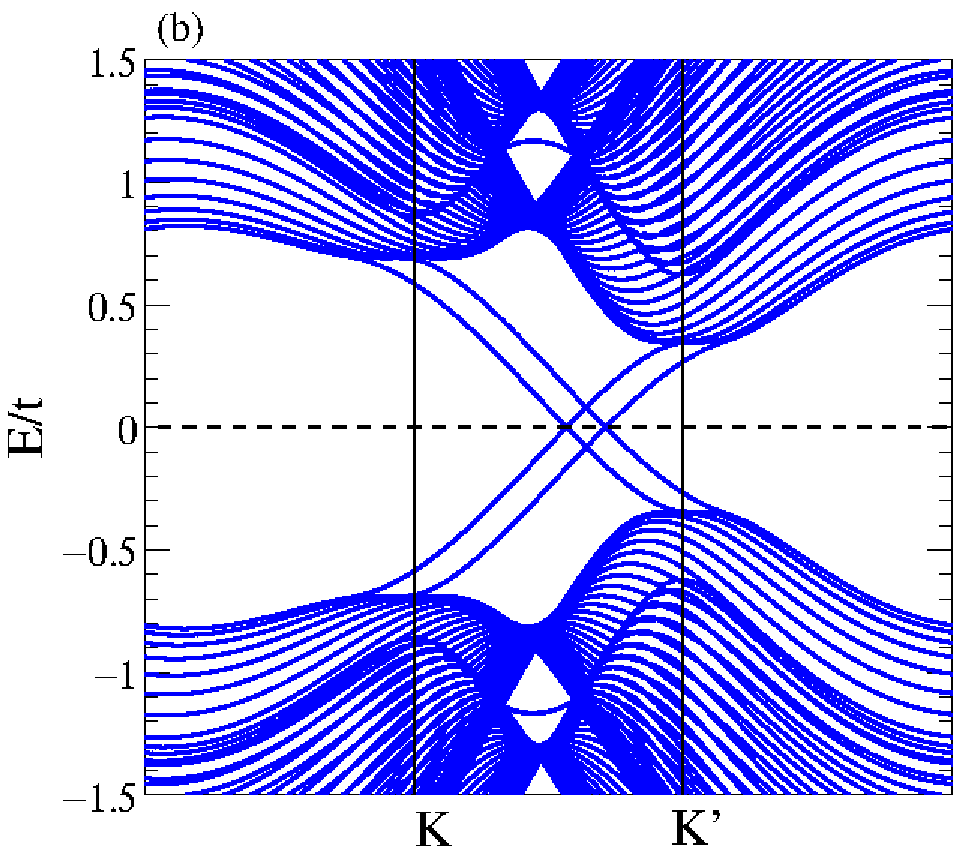}&\includegraphics[width=0.33\columnwidth]{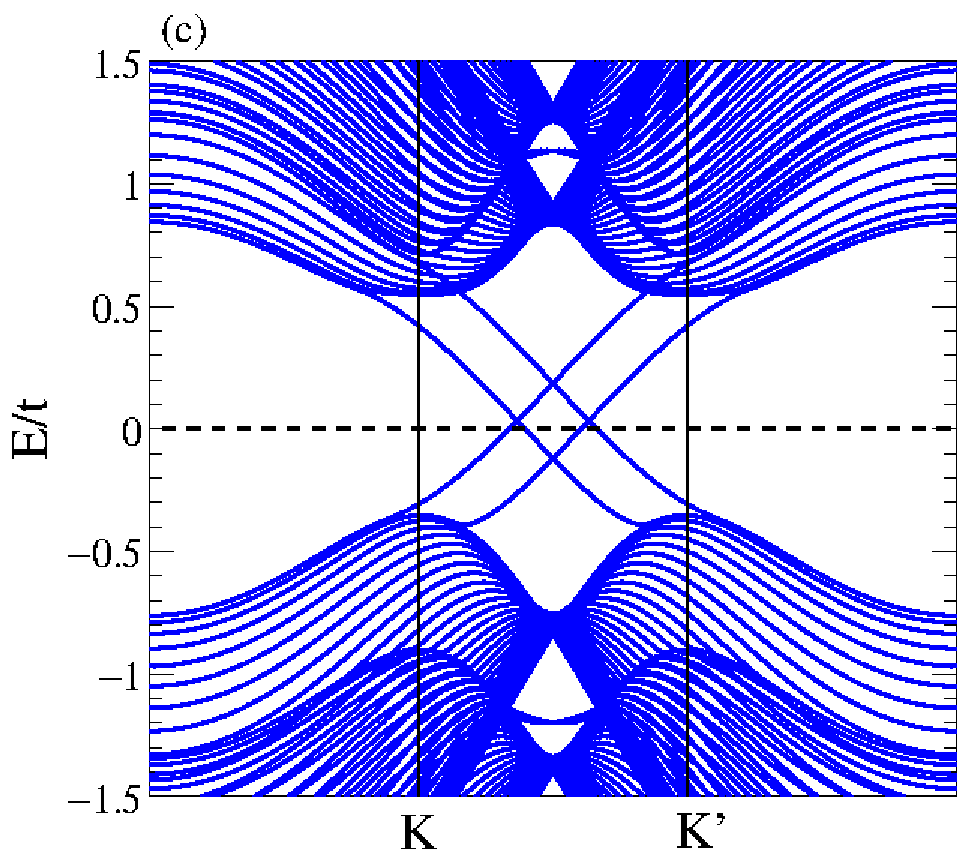}\\
\includegraphics[width=0.33\columnwidth]{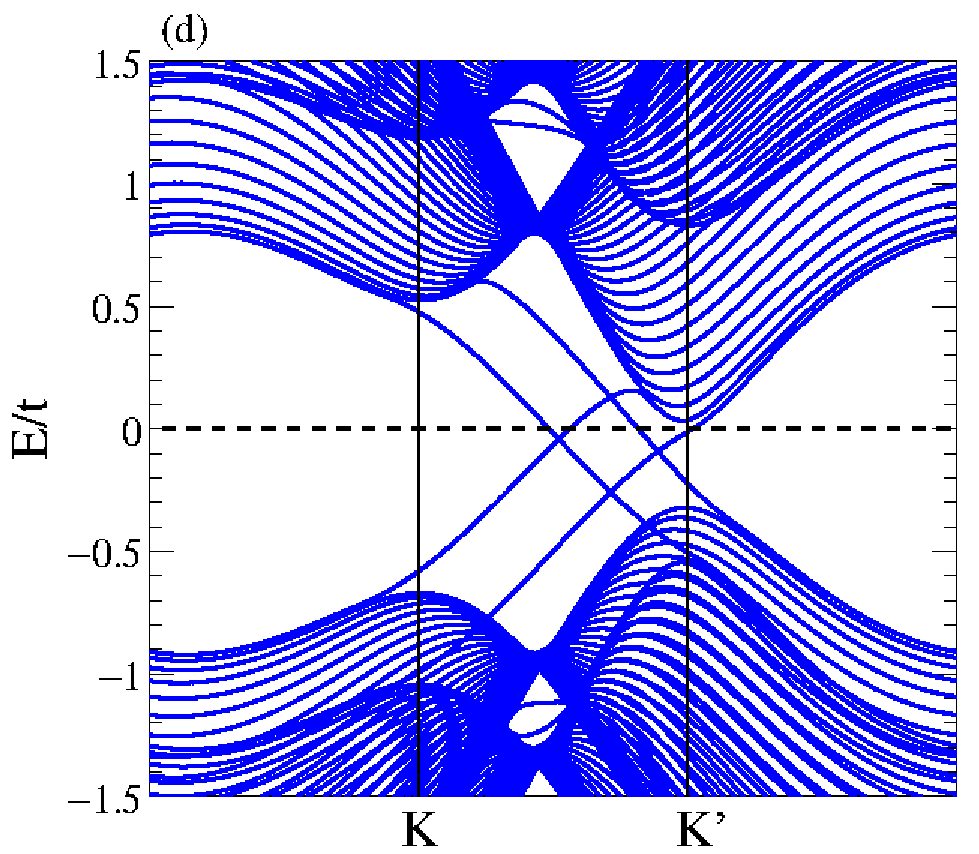}
\includegraphics[width=0.33\columnwidth]{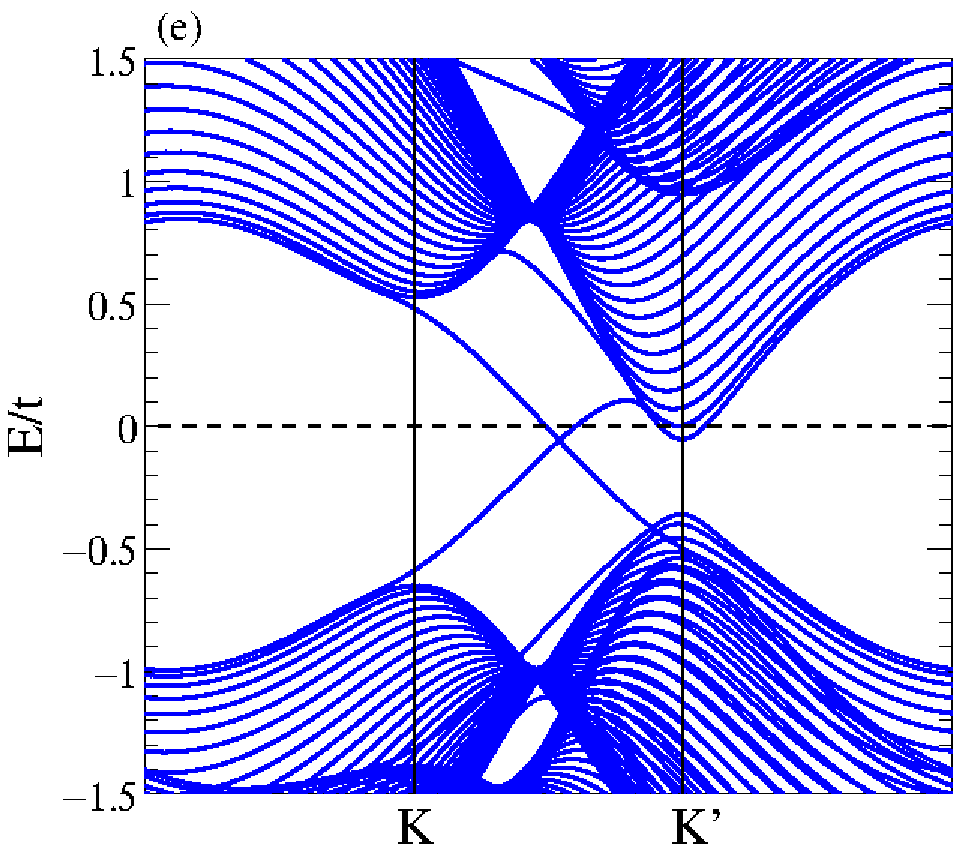}&
\includegraphics[width=0.33\columnwidth]{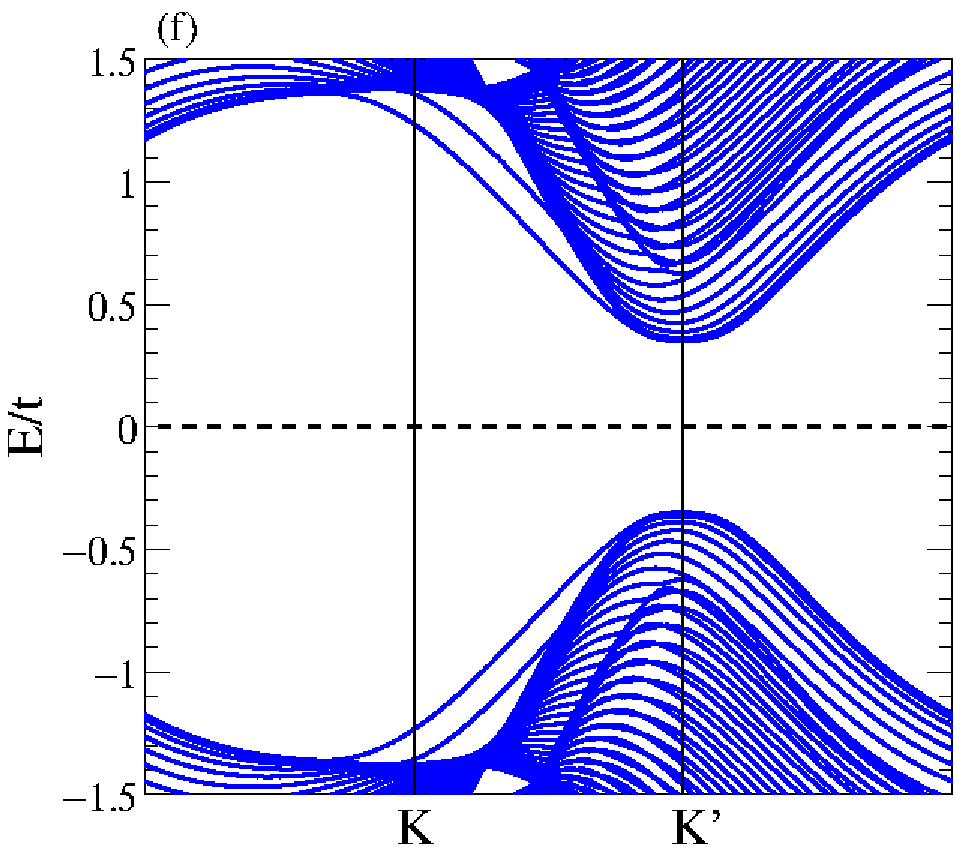}&
\end{array}
$
\caption{Electronic band structure of an AB bilayer HM on zigzag
nanoribbons of a width $W = 60$ atoms. Calculations are done for $t_2 = 0.1t$, $\Phi_1=\Phi_2= \frac{\pi}2$, $t_{\perp}=0.5t$ and for (a) $M_1=M_2=0$, (b) $M_1=M_2=\sqrt{3}t_2$, (c) $M_1=-M_2=\sqrt{3}t_2$,
(d) $M_1=0, M_2=3\sqrt{3}t_2$, (e) $M_1=0, M_2=5\sqrt{3}t_2$, (f) $M_1=5\sqrt{3}t_2, M_2=5\sqrt{3}t_2$.}
\label{band-HM-mass}
\end{figure}

As shown in figure ~\ref{band-HM-mass}, the $C=\pm 2$ Chern insulating phase occurs as far as the Semenoff mass $|M_l|<|M_{lc}|$, where
\begin{eqnarray}
M_{lc}=3\sqrt{3}t_2\sin \Phi_l.
\label{Mlc}
\end{eqnarray}
$M_{lc}$ is the critical mass at which the transition from a topological phase ($C_l=\pm 1$) to a trivial gapped phase ($C_l=0$), takes place in the monolayer HM~\cite{Haldane}.

\begin{figure}[hpbt] 
\begin{center}
$
\begin{array}{cc}
\includegraphics[width=0.5\columnwidth]{HM-AB-ZZ-b.eps}
\includegraphics[width=0.5\columnwidth]{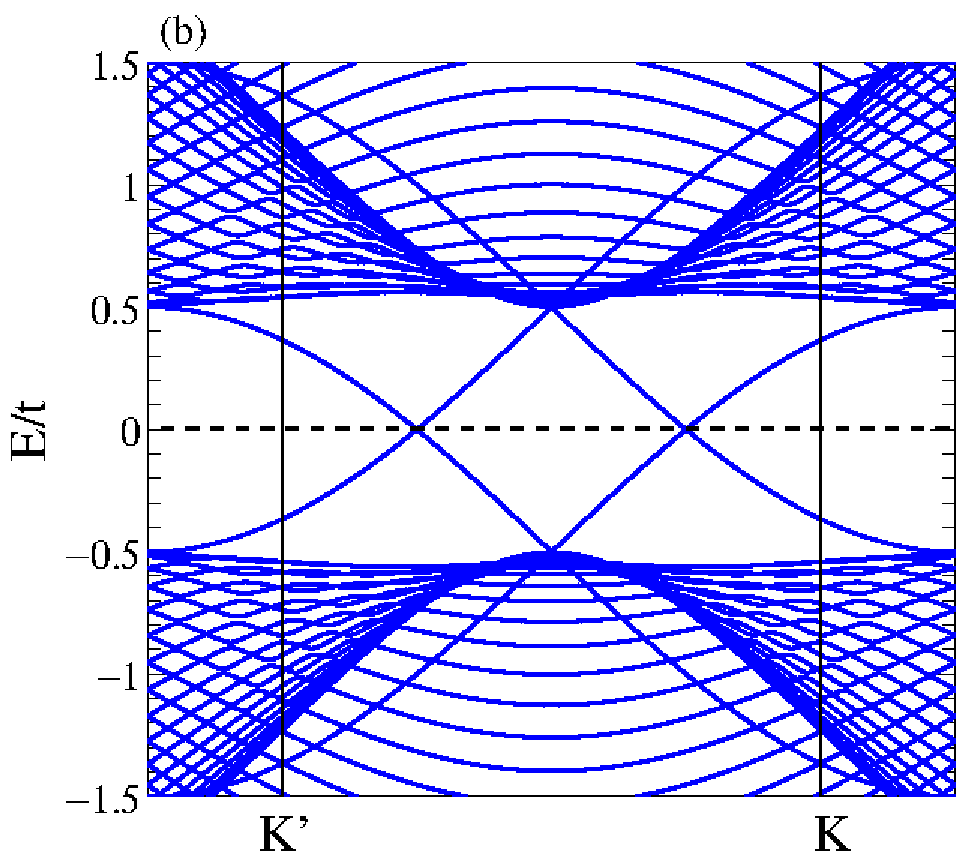}
\end{array}
$
\end{center}
\caption{Electronic band structure of an AB bilayer HM on (a) zigzag and (b) armchair nanoribbons of a width $W = 60$ atoms. 
Calculations are done for $t_2 = 0.1t$, $t_{\perp} = 0.5t$, $\Phi_1=\Phi_2=\frac{\pi}2$ and
$M_1=M_2=0$.}
\label{band-HM-AC}
\end{figure}

This behavior is independent of the nature (zigzag or armchair) of the ribbon boundaries as shown in Fig.~\ref{band-HM-AC}.

\begin{figure}[hpbt] 
\begin{center}
$
\begin{array}{cc}
\includegraphics[width=0.5\columnwidth]{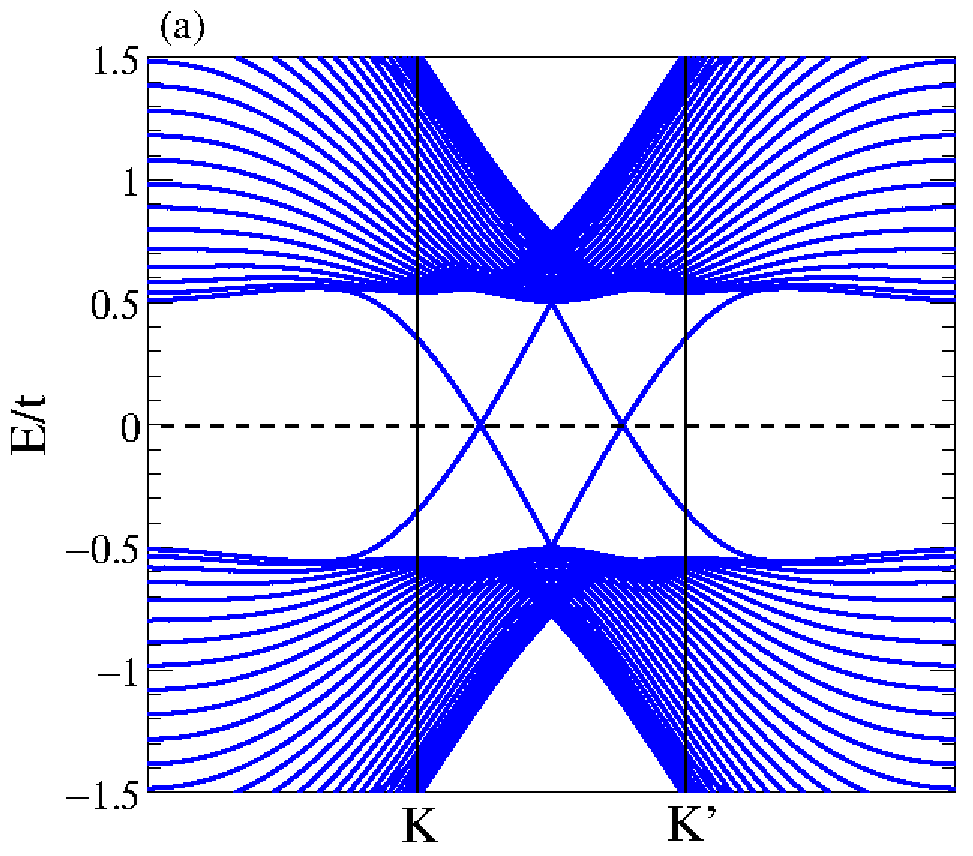}
\includegraphics[width=0.5\columnwidth]{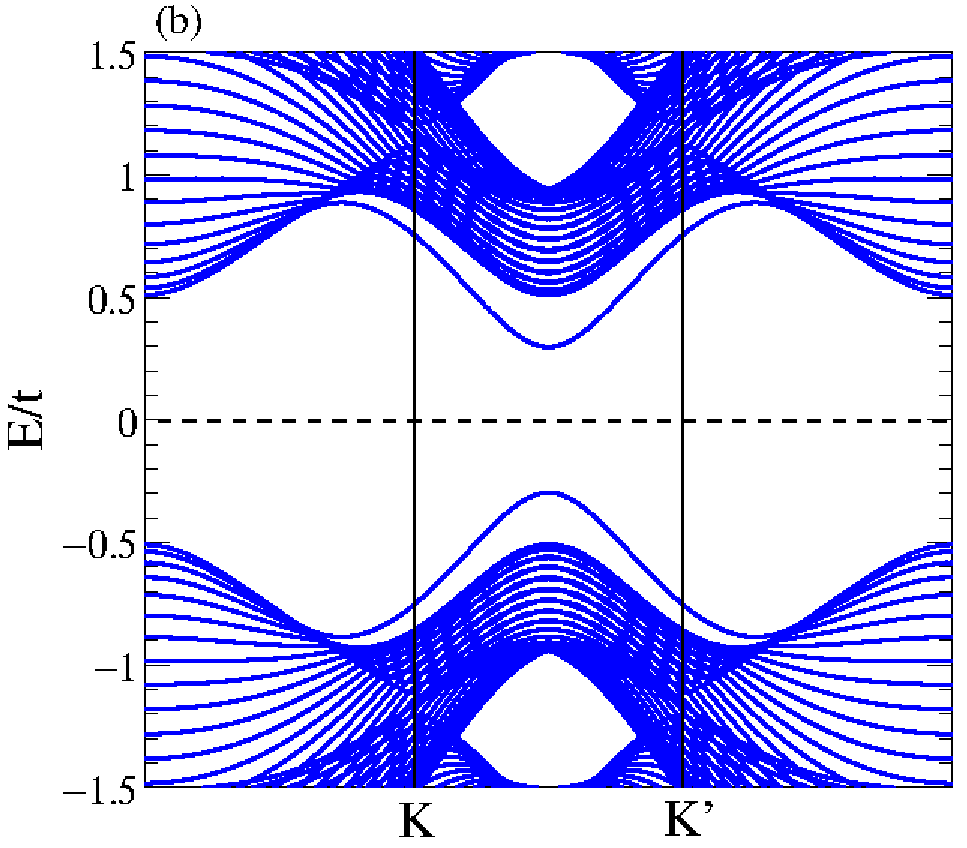}\\
\includegraphics[width=0.5\columnwidth]{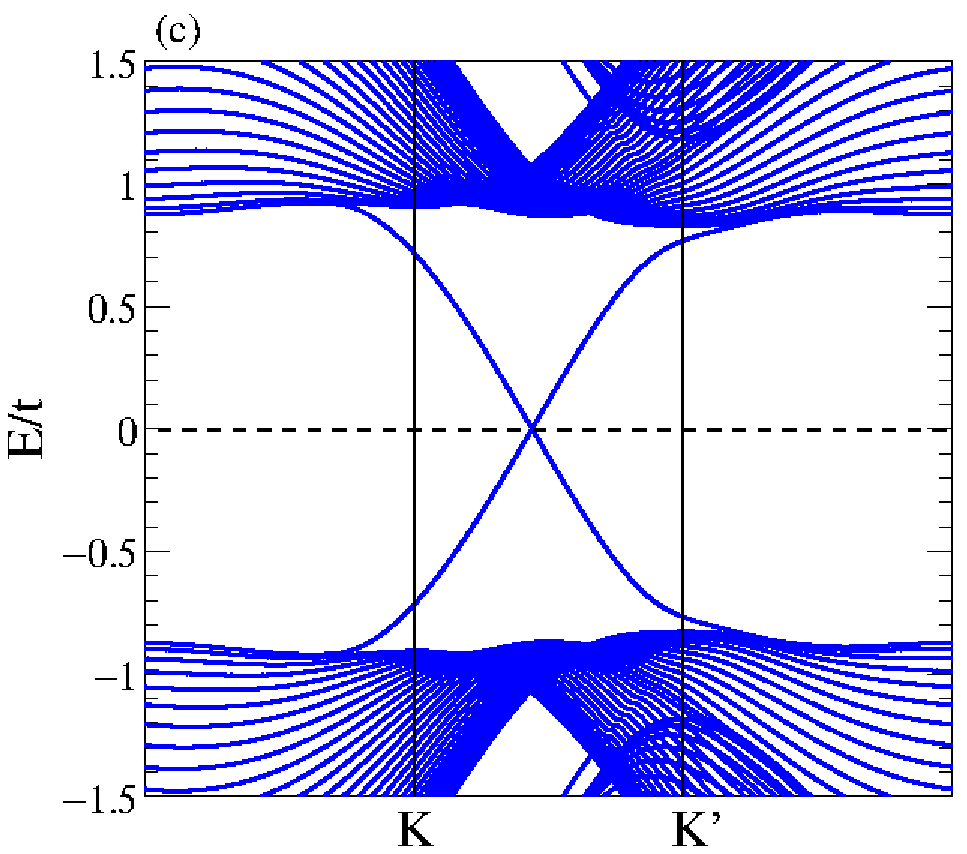}
\includegraphics[width=0.5\columnwidth]{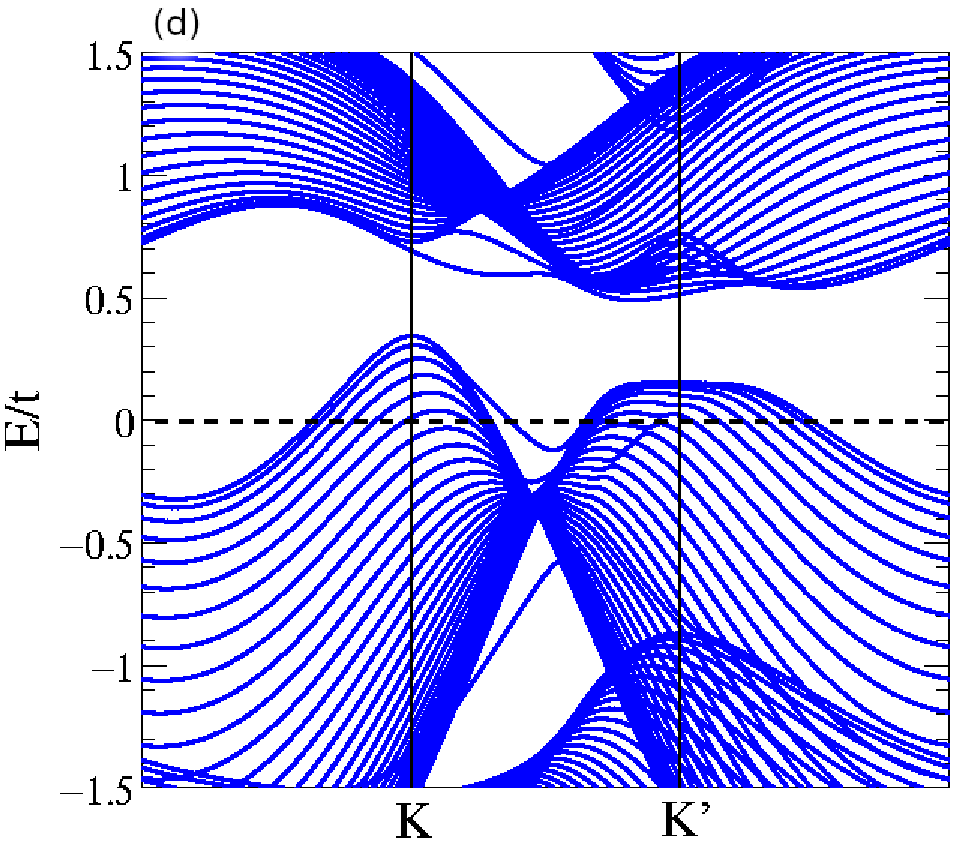}
\end{array}
$
\end{center}
\caption{Electronic band structure of an AA Bilayer HM on zigzag
nanoribbons of a width $W = 60$ atoms for $t_{\perp} = 0.5t$,
(a) $\Phi_1=\Phi_2=\frac{\pi}2$, $M_1=M_2=0$, (b) $\Phi_1=\Phi_2=-\frac{\pi}2$, $M_1=M_2=0$, (c) $\Phi_1=\Phi_2=\frac{\pi}2$, $M_1=0$, $M_2=5\sqrt{3}t_2$ and
(d) $\Phi_1=\frac{\pi}2, \Phi_2=0$,  $M_1=\sqrt{3}t_2$, $M_2=0$. Calculations are done for
$t_2=0.1t$ in (a), (b) and (d) and $t_2=0.2t$ in (c).}
\label{band-HM-AA}
\end{figure}

Regardless of the stacking type (AB or AA), the bilayer HM is~\cite{Dutta}:
$(i)$ a trivial insulator, if the layers have opposite Chern numbers $C_1=-C_2$,
$(ii)$ a topological chiral insulator with $C=\pm 2$, if the layers have the same chirality ($C_1=C_2$),
$(iii)$ and a Chern insulator with $C=\pm 1$ if one layer has a non-vanishing Chern number $C_1=\pm 1$ and the other layer is a trivial insulator $C_2=0$, as depicted in Fig.~\ref{band-HM-AA} showing the band structure of an AA bilayer HM on zigzag ribbons.

\section{B. Modified Haldane model in AA bilayer}
\label{App-mHM-AA}
\renewcommand{\thefigure}{B\arabic{figure}}
\setcounter{figure}{0}

\begin{figure}[hpbt] 
$
\begin{array}{cc}
\includegraphics[width=0.5\columnwidth]{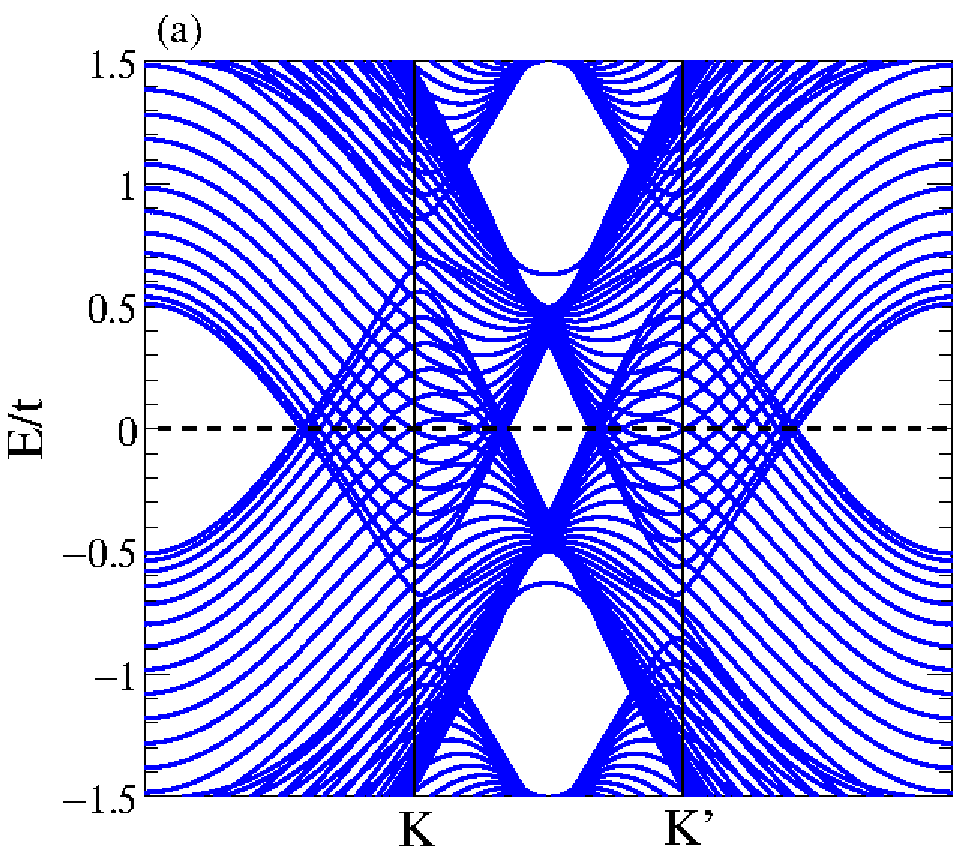}
\includegraphics[width=0.5\columnwidth]{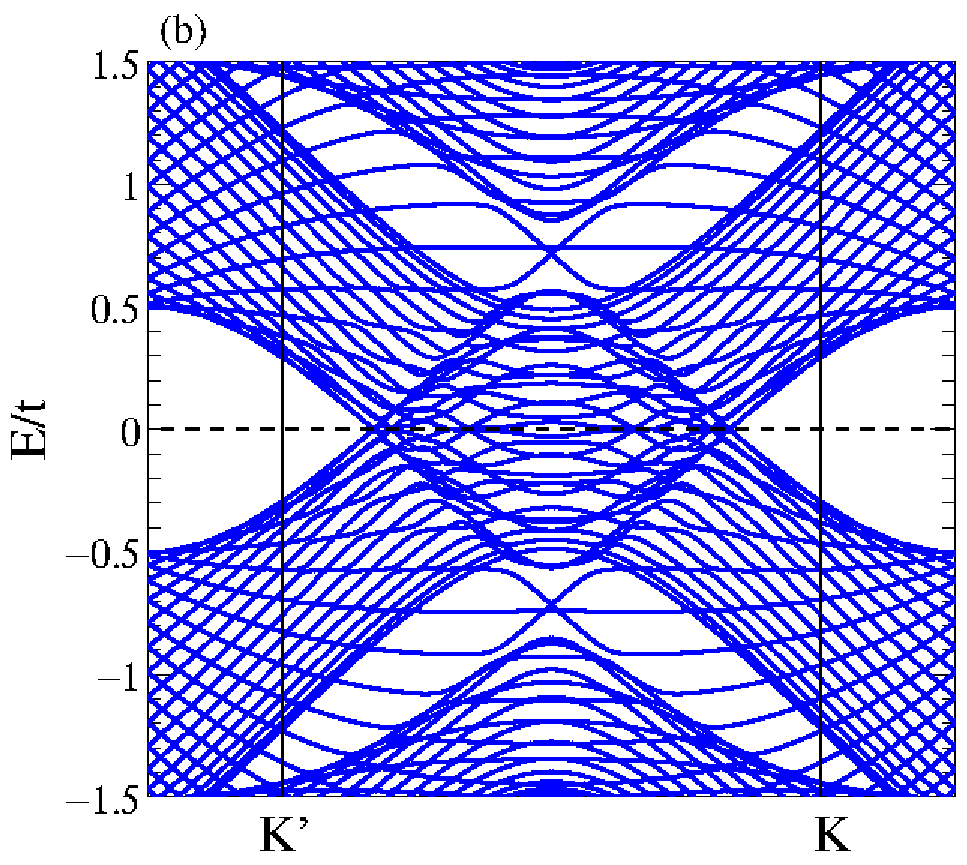}\\
\includegraphics[width=0.5\columnwidth]{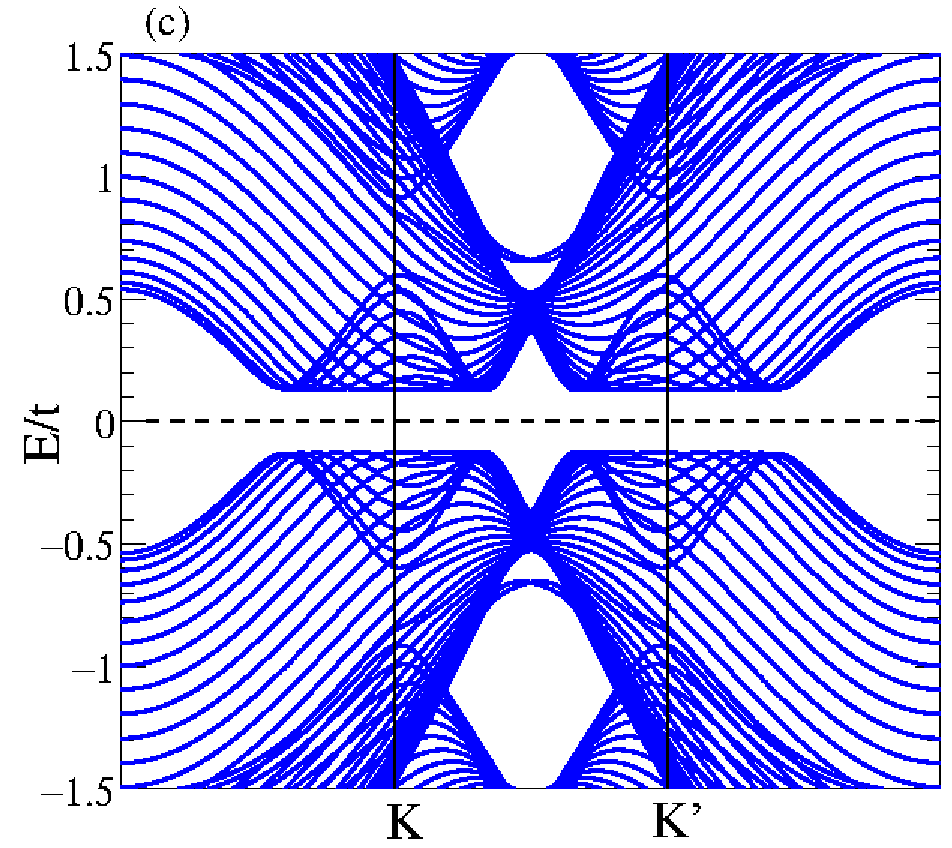}
\includegraphics[width=0.5\columnwidth]{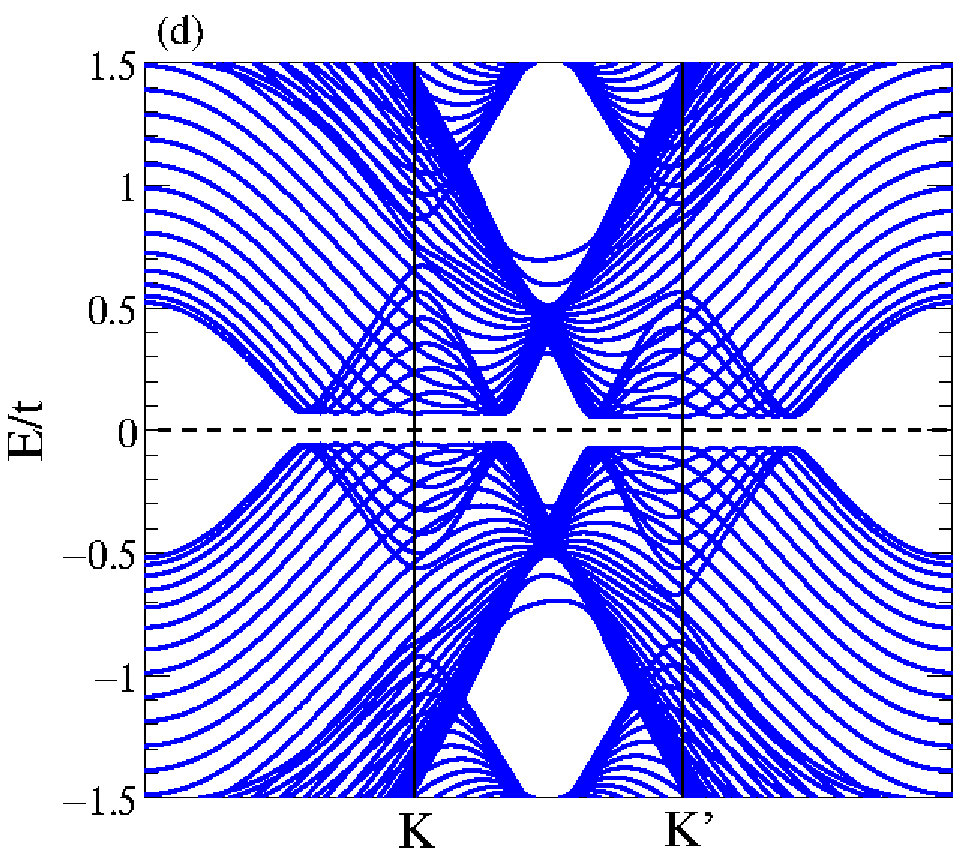}
\end{array}
$
\begin{center}
\end{center}
\caption{Band structure of the mHM on AA nanoribbons of a width of $W=60$ atoms with  (a,c,d) zigzag and (b) armchair boundaries. Calculations are done for $t_2 = 0.1t$, $t_{\perp}=0.5t$, (a) and (b) $\Phi_1=-\Phi_2= \frac{\pi}2$, $M_1=M_2=0$, while in (c)  $\Phi_1=-\Phi_2= \frac{\pi}2$, $M_1=-M_2=\sqrt{3}t_2$ and (d) $\Phi_1=-\Phi_2= \frac{\pi}2$, $M_1=0$, $M_2=\sqrt{3}t_2$. }
\label{AA}
\end{figure}
Figure~\ref{AA} shows the band structure of the mHM in AA stacked ribbons with zigzag and armchair boundaries in the case of opposite complex phases $\Phi_1=-\Phi_2$.
In the absence of the Semenoff masses ($M_1=M_2=0$), the system remains gapless under the interlayer coupling. However, it turns to a trivial insulator if the layers have Semenoff mass terms. \

Therefore, in the absence of the Semenoff masses, the Fermi surface (Fig.~\ref{Fig-intro}) of the mHM in AA stacked bilayer is,
contrary to the AB stacking, stable against the interlayer hopping which cannot induce a gap opening.\

To understand the Fermi surface stability, we start by writing the corresponding Hamiltonian in the basis of the four orbitals forming the unit cell ($A_1,B_1,A_2,B_2$) and we consider, for simplicity, the case of opposite complex NNN phases $\Phi_1=-\Phi_2=\frac{\pi}2$ to have a vanishing global energy shift ($a^0_{\mathbf{k}}=0$, Eq.~\ref{al} of the main text)
\begin{eqnarray}
 H_{AA-mHM}(\mathbf{k})=
\begin{pmatrix}
a_{\mathbf{k}} & f_{\mathbf{k}} &2t_{\perp}&0 \\
f^{\ast}_{\mathbf{k}} & a_{\mathbf{k}} &0&2t_{\perp}\\
2t_{\perp}&0&-a_{\mathbf{k}} & f_{\mathbf{k}}\\
0&2t_{\perp}&f^{\ast}_{\mathbf{k}} & -a_{\mathbf{k}}
\end{pmatrix}.
\label{AA-mHM}
\end{eqnarray}
This Hamiltonian can be written, using the sublattice and the layer pseudospin matrices $\boldsymbol{\sigma}$ and $\boldsymbol{\tau}$, as
\begin{eqnarray}
H_\text{AA-mHM}(\mathbf{k})&=&\left( b_{\mathbf{k}}\sigma_x+c_{\mathbf{k}}\sigma_y\right)\tau_0
+2t_{\perp}\sigma_0\tau_x+ a_{\mathbf{k}}\sigma_0\tau_z,\nonumber\\
\label{HBLAA}
\end{eqnarray}
where $a_{\mathbf{k}}$ is given by Eq.~\ref{al} in the main text.\

The Hamiltonian of Eq.~\ref{HBLAA} breaks TRS, $\mathcal{T}=K\tau_x$, the charge conjugation, represented by $\mathcal{C}=\sigma_z\tau_zK$ with $\mathcal{C}^2=\mathds{1}$, and the chirality $\mathcal{S}=\tau_z\sigma_z$.\

%However, as for the HM in AA bilayer, $H_\text{AA-mHM}$ (Eq.~\ref{HBLAA}) is invariant under the TRS operator represented by $\mathcal{T_x}=K\tau_x$ since $\tau_x H_\text{AA-mHM}^{\ast}(-\mathbf{k})=H_\text{AA-mHM}(\mathbf{k})$.
The gap separating the two bands, $E_{-,-}(\mathbf{k})$ and $E_{+,-}(\mathbf{k})$, around the zero energy is $\Delta=\mathrm{min}_{\mathbf{k}}\left(\Delta_{\mathbf{k}}\right)$, where 
\begin{eqnarray}
 &&\Delta_{\mathbf{k}}=2\sqrt{A_{\mathbf{k}}-B_{\mathbf{k}}},\;
 A_{\mathbf{k}}=a^2_{\mathbf{k}}+|f_{\mathbf{k}}|^2+4t^2_{\perp},\nonumber\\ &&B_{\mathbf{k}}=2|f_{\mathbf{k}}|\sqrt{a^2_{\mathbf{k}}+4t^2_{\perp}}
\end{eqnarray}
$\Delta_{\mathbf{k}}=0$ leads to
\begin{eqnarray}
|f_{\mathbf{k}}|^2=a^2_{\mathbf{k}}+4t^2_{\perp},
\label{FL}
\end{eqnarray}
which defines a closed Fermi line. \

For $a_{\mathbf{k}}=0$, Eq.~\ref{FL} corresponds to the Fermi line of the AA graphene bilayer in the absence of NNN hopping terms. \

For $t_{\perp}=0$, Eq.~\ref{FL} describes the mHM in AA bilayer with a particle-hole Fermi line obeying to $|f_{\mathbf{k}}|=|a_{\mathbf{k}}|$.\

By turning on $t_{\perp}$, this Fermi line is, simply, shifted but cannot be gapped (Eq.~\ref{FL}).
The mHM on AA bilayer remains, then, metallic for vanishing Semenoff masses.\

\section{C. Phase transition from AB to BA bilayer of the modified Haldane model}
\label{App-AB-BA}
\renewcommand{\thefigure}{C\arabic{figure}}
\setcounter{figure}{0}

We consider a generic Hamiltonian that, continuously, interpolates between AB $(\theta=0)$ and BA $(\theta=\pi/2)$ stackings of two single graphene layers described by the mHM. The interlayer hopping term, for a given value of $\theta$ between these limits, does not represent physical coupling.
We take, for simplicity, vanishing Semenoff masses and inplane complex phases $\Phi_1=-\Phi_2=\frac{\pi}2$. The Bloch Hamiltonian written in the basis ($A_1,B_1,A_2,B_2$) is
\begin{eqnarray}
 H_{B}(\mathbf{k},\theta)=
\begin{pmatrix}
a_{\mathbf{k}}& f_{\mathbf{k}} &0& 2t_{\perp} c_{\theta}\\
f^{\ast}_{\mathbf{k}} & a_{\mathbf{k}} & 2t_{\perp} s_{\theta} &0\\
0&2t_{\perp} s_{\theta}&-a_{\mathbf{k}}& f_{\mathbf{k}}\\
2t_{\perp} c_{\theta}&0&f^{\ast}_{\mathbf{k}} & -a_{\mathbf{k}}
\end{pmatrix}
\label{HBLtheta}
\end{eqnarray}
where $f_{\mathbf{k}}=t\sum_{i=1}^3 e^{i\mathbf{k}\cdot\boldsymbol{\delta}_i}$ and $a_{\mathbf{k}}=-2t_2\sum_{i=1}^3  \sin\left(\mathbf{k}\cdot\mathbf{a}_i\right)$. The AB (BA) stacking corresponds to $c_{\theta}\equiv\cos\theta=1$ and $s_{\theta}\equiv\sin\theta=0$ ($c_{\theta}=0$, $s_{\theta}=1$). \

By varying $\theta$, the system can be interpolated between the two stacking configurations, without going through the AA stacking.\
Using the intralayer and interlayer pseudo-spin matrices $\boldsymbol{\sigma}$ and $\boldsymbol{\tau}$, the Bloch Hamiltonian of Eq.~\ref{HBLtheta} becomes
\begin{eqnarray}
H_{B}(\mathbf{k},\theta)&=&\left( b_{\mathbf{k}}\sigma_x+c_{\mathbf{k}}\sigma_y\right)\tau_0
+2t_{\perp}\left[\left(c_{\theta}\sigma_++s_{\theta}\sigma_-\right)\tau_+\right.\nonumber\\
&+&\left.2t_{\perp}\left(c_{\theta}\sigma_-+s_{\theta}\sigma_+\right)\tau_-\right],
\label{HBL2theta}
\end{eqnarray}
where $b_{\mathbf{k}}$, $c_{\mathbf{k}}$, $\sigma_{\pm}$ and $\tau_{\pm}$ are given in the main text (Eqs.~\ref{HMmL} and \ref{HBL2}).\

The Hamiltonian of Eq.~\ref{HBL2theta} breaks TRS, chirality but is invariant under charge-conjugation and inversion since $M_l=0$ and $\Phi_1=-\Phi_2=\frac{\pi}2$.\
The corresponding energy spectrum is expressed, as in Eq.~\ref{E+-} of the main text with
\begin{eqnarray}
 A_{\mathbf{k}}&=&a^2_{\mathbf{k}}+|f_{\mathbf{k}}|^2+2t_{\perp}^2,\nonumber\\
 B_{\mathbf{k}}&=&2\sqrt{|f_{\mathbf{k}}|^2\left(a^2_{\mathbf{k}}+t_{\perp}^2\right)+c^2_{2\theta}t_{\perp}^4+s_{2\theta}t_{\perp}^2\left( b^2_{\mathbf{k}}-c^2_{\mathbf{k}}\right)}\nonumber\\
 \label{ABktheta}
\end{eqnarray}
The gap separating the lowest energy band around zero energy is $\Delta=\min_{\mathbf{k}}\left(\Delta_{\mathbf{k}}\right)=2\sqrt{A_{\mathbf{k}}-B_{\mathbf{k}}}$. \

$\Delta_{\mathbf{k}}$ closes, for uncoupled layers ($t_{\perp}=0$), for $A_{\mathbf{k}}=B_{\mathbf{k}}$ which defines, as we have seen in the main text, two non-intersecting closed Fermi lines.
For a non vanishing interlayer hopping $t_{\perp}$, $\Delta_{\mathbf{k}}$ can be expressed as

\begin{eqnarray}
\Delta_{\mathbf{k}}=2\frac{\sqrt{A^2_{\mathbf{k}}-B^2_{\mathbf{k}}}}{A_{\mathbf{k}}+B_{\mathbf{k}}}
\sim \frac{2t_{\perp}}{|f_{\mathbf{k}}|}\sqrt{\left(1-s_{2\theta}\right)b^2_{\mathbf{k}}+\left(\lambda^2_{\mathbf{k}}-c^2_{2\theta} \right)t_{\perp}^2},\nonumber\\
\label{Delta}
\end{eqnarray} 
where we introduced the parameter $\lambda_{\mathbf{k}}$ defined as 
\begin{eqnarray}
a^2_{\mathbf{k}}=|f_{\mathbf{k}}|^2+\left(2\lambda_{\mathbf{k}}-1\right)t_{\perp}^2
\label{lambda}
\end{eqnarray}
In Eq.~\ref{Delta}, the numerator is given by its exact expression, whereas the denominator has been approximated to the zeroth order in $t_{\perp}$.

According to Eq.~\ref{Delta}, $\Delta_{\mathbf{k}}$ cannot vanish for $\theta\neq \pi/4,3\pi/4$. \

For $\theta=\pi/4 $, $\Delta_{\mathbf{k}}=0$ if $\lambda_{\mathbf{k}}=0$, which give rise, according to Eq.~\ref{lambda}, to a closed loop defined by $a^2_{\mathbf{k}}+t_{\perp}^2=|f_{\mathbf{k}}|^2$.\

For $\theta=3\pi/4$, the closing of the gap $\Delta_{\mathbf{k}}$ requiers $\lambda_{\mathbf{k}}=0$ and $b_{\mathbf{k}}=0$.\

The critical values $\theta=\pi/4,3\pi/4 $ correspond to the semimetallic phase marking the transition from the topological Chern insulator phase $C=2$, occurring for $0\le\theta<\pi/4 $ to the Chern insulator phase $C=-2$, associated to $3\pi/4<\theta\le\pi $.
At this topological phase transition, the gap closes at the four Dirac points of the bilayer system where, right after the transition, the signs of the Dirac masses flip (see Eq.~\ref{Chern-def}), inducing a variation $\Delta C=-4$ of the Chern number.

The flipping of the Chern number sign, at the crossing form the AB to the BA stacking, could be understood from Fig.~\ref{AB-AA} of the main text. Such crossing can be regarded as a sign change of the complex phases: since the inplane sublattices have opposite fluxes, the AB stacking corresponds to the dimer ($A_1$,$B_2$) for which the complex phases are $\Phi_1=-\Phi_2=\pi/2$ while the BA stacking is ascribed to the dimer ($B_1$,$A_2$) with $\Phi_1=-\Phi_2=-\pi/2$ (Fig.~\ref{AB-AA} of the main text).

\section{D. Effective two-band model for the modified Haldane model in AB bilayer}
\label{App-mHM-model}
\renewcommand{\thefigure}{D\arabic{figure}}
\setcounter{figure}{0}

To derive the low energy Hamiltonian given by Eq.~\ref{Heff} in the main text, we use the 
L\"owdin partitioning method~\cite{Lowdin,McCann} in the case of bilayer graphene. 
For simplicity, we consider the case $\Phi_1=-\Phi_2=\pm\frac{\pi}2$ to remove the energy-shift terms $a^0_{l,\mathbf{k}}$ (Eq.\ref{al}). We rewrite the full Hamiltonian, (Eq.~\ref{HBL} of the main text) in the basis ($A_2,B_1,A_1,B_2$) as
\begin{eqnarray}
 H_{B}(\mathbf{k})=
\begin{pmatrix}
H_{\alpha\alpha} & H_{\alpha\beta} \\
H_{\beta\alpha} & H_{\beta\beta}
\end{pmatrix},
\end{eqnarray}
where
\begin{eqnarray}
&&H_{\alpha\alpha}=
\begin{pmatrix}
 -a_{\mathbf{k}}+M_2& 0 \\
0 & a_{\mathbf{k}}-M_1
\end{pmatrix},\nonumber\\
&&H_{\alpha\beta}=H_{\beta\alpha}
\begin{pmatrix}
0&f_{\mathbf{k}} \\
f^{\ast}_{\mathbf{k}} & 0
\end{pmatrix},\,
H_{\beta\beta}=
\begin{pmatrix}
 a_{\mathbf{k}}+M_1& t_{\perp} \\
 t_{\perp}& -a_{\mathbf{k}}-M_2
\end{pmatrix}.\nonumber\\
\end{eqnarray}
The corresponding effective Hamiltonian is~\cite{McCann}

\begin{eqnarray}
 H_\text{eff}({\mathbf{k}},E)=H_{\alpha\alpha}+H_{\alpha\beta}\left( E-H_{\beta\beta}\right)^{-1} H_{\beta\alpha},
\end{eqnarray}
which reduces in the limit $M_l\sim t_2\ll t_{\perp}$ ($l=1,2$), and for $E\sim 0$ to
\begin{eqnarray}
H_\text{eff}({\mathbf{k}},E=0)\equiv H_\text{eff}({\mathbf{k}})\sim H_{\alpha\alpha}-\frac 1{X^2}H_{\alpha\beta}H_{\beta\beta} H_{\beta\alpha},\nonumber\\
\end{eqnarray}
where $X^2=\left(a_{\mathbf{k}}+M_1\right)\left(a_{\mathbf{k}}+M_2\right)+4t^2_{\perp}$.\

Assuming $M_l\frac{|f_{\mathbf{k}}|^2}{X^2}\ll M_{l^{\prime}}\sim t_2 \, (l,l^{\prime}=1,2)$, the corresponding effective Hamiltonian gives rise to Eq.~\ref{Heff} of the main text.\
\section{E. Effect of complex phases and of Semenoff masses on AB bilayer of modified Haldane model}
\label{App-mHM-phi-mass}
\renewcommand{\thefigure}{E\arabic{figure}}
\setcounter{figure}{0}
It is noteworthy that the  induced Chern insulator in the AB bilayer mHM occurs as far as $\Phi_1$ and $\Phi_2$ have opposite signs and not only in the case where $\Phi_1=-\Phi_2=\frac{\pi}2$, which we considered to obtain simple analytical expressions. This feature is shown in Fig.~\ref{phi}.
The effect of the Semenoff masses on the mHM in monolayer graphene nanoribbon is represented in Fig.~\ref{mass-ML} showing that the mass term lifts the degeneracy of the antichiral edge modes which
survive as far as $M<M_c\equiv 3\sqrt{3}t_2\sin \Phi$.\

In Fig.~\ref{mass}, we plot the band structure of the mHM on AB bilayer honeycomb lattices with different choices of intralayer Semenoff masses.\
Figure \ref{mass} shows that, in AB bilayer mHM, the chiral edge states emerge as far as $M_l<M_{lc}$ (Eq.~\ref{Mlc}), for which the nodal lines, originating from the overlap of the two layer bands, can occur, as discussed in the main text (Fig.~\ref{Fig-intro}).\\
\begin{widetext}
\begin{figure*}[hpbt]
\begin{center}
$\begin{array}{c}
\includegraphics[width=0.4\columnwidth]{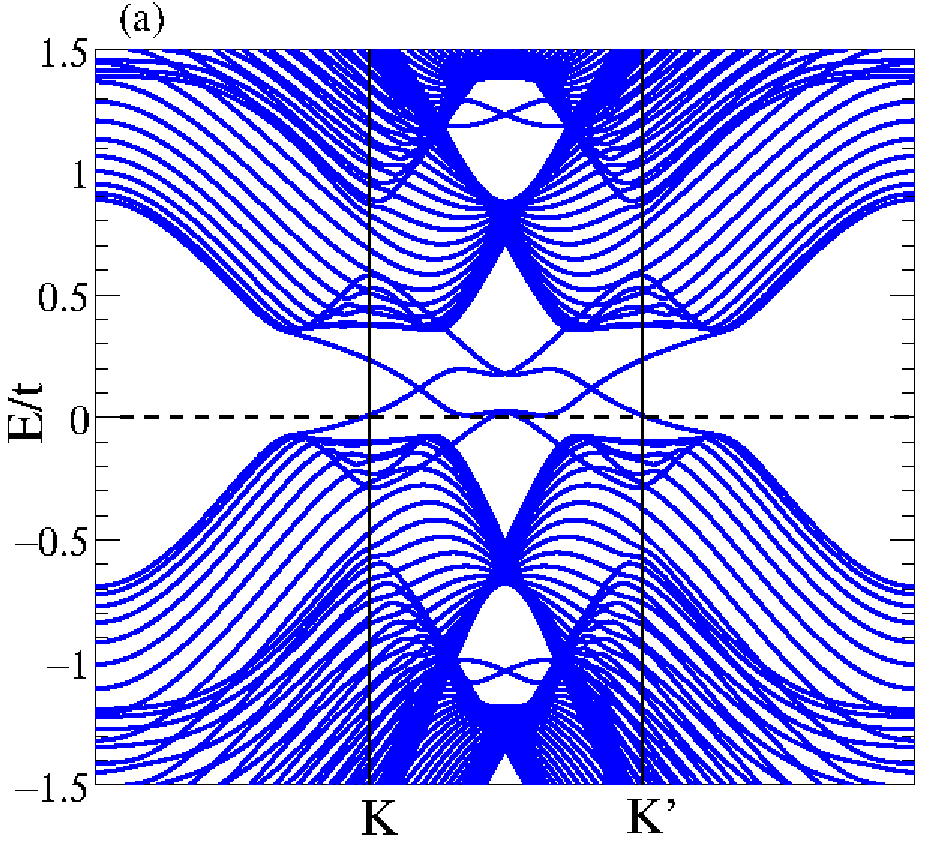}
\includegraphics[width=0.4\columnwidth]{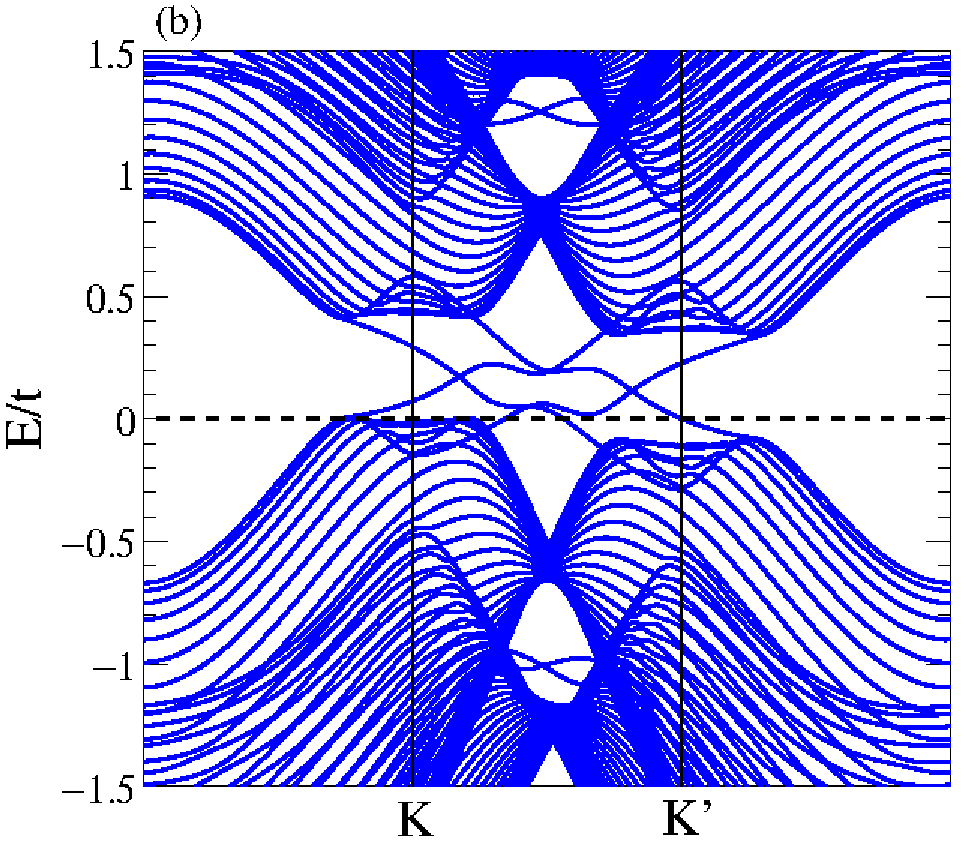}
\includegraphics[width=0.4\columnwidth]{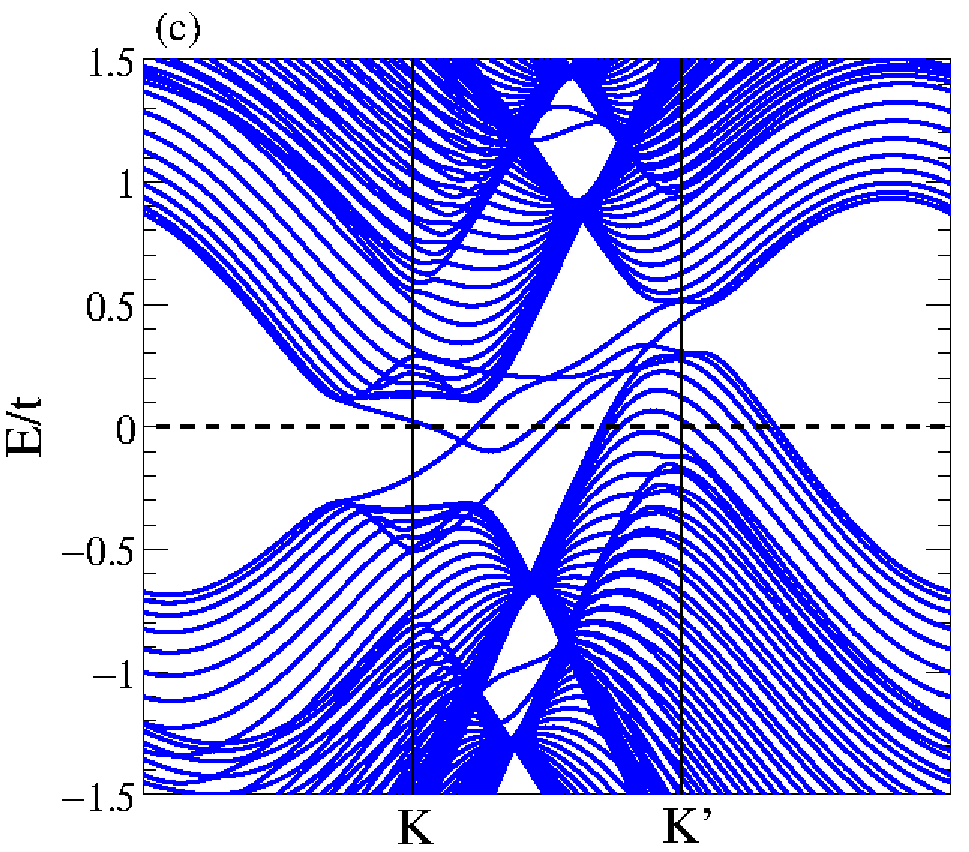}
\end{array}$
\end{center}
\caption{Band structure of the mHM in AB stacked bilayer ribbons of a width $W=60$ atoms, for $t_2=0.1t$, $t_{\perp}=0.5 t$, $M_1=M_2=0$ and (a) $\Phi_1=-\Phi_2=\frac{\pi}3$, $\Phi_1=\frac{\pi}3$, $\Phi_2=-\frac{\pi}4$ and (c) $\Phi_1=0,\Phi_2=-\frac{\pi}2$.}
\label{phi}
\end{figure*} 
\begin{figure*}[hpbt]
\centering
$\begin{array}{cccc}
\includegraphics[width=0.4\columnwidth]{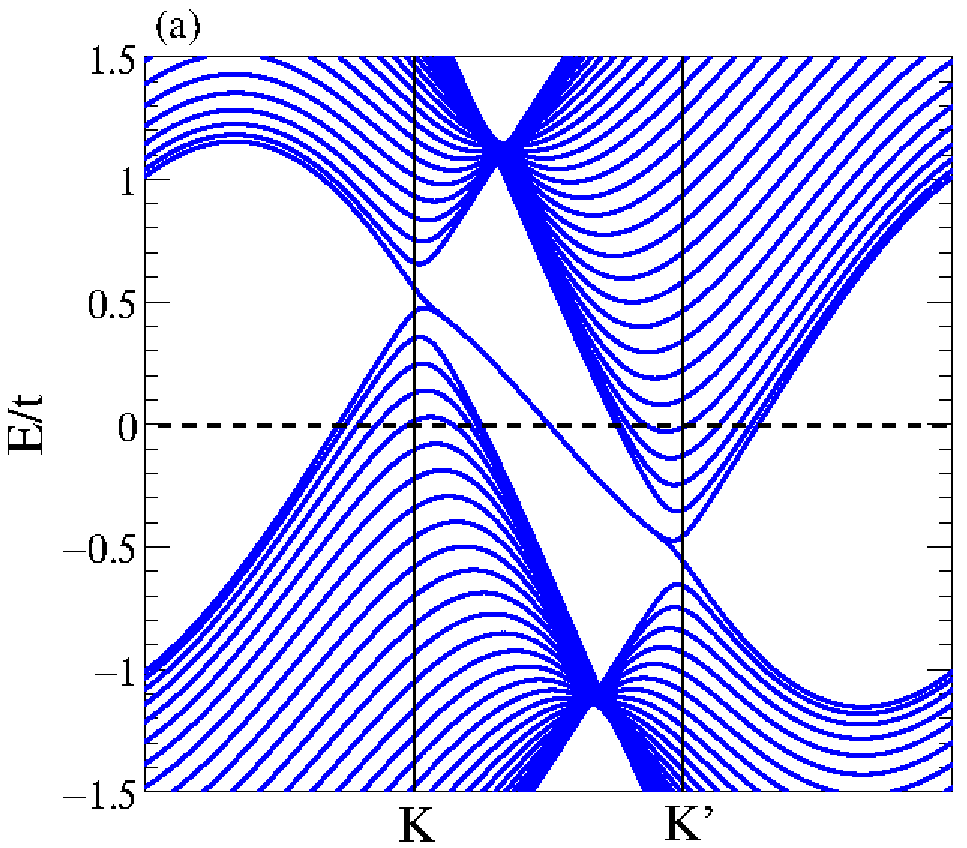}
\includegraphics[width=0.4\columnwidth]{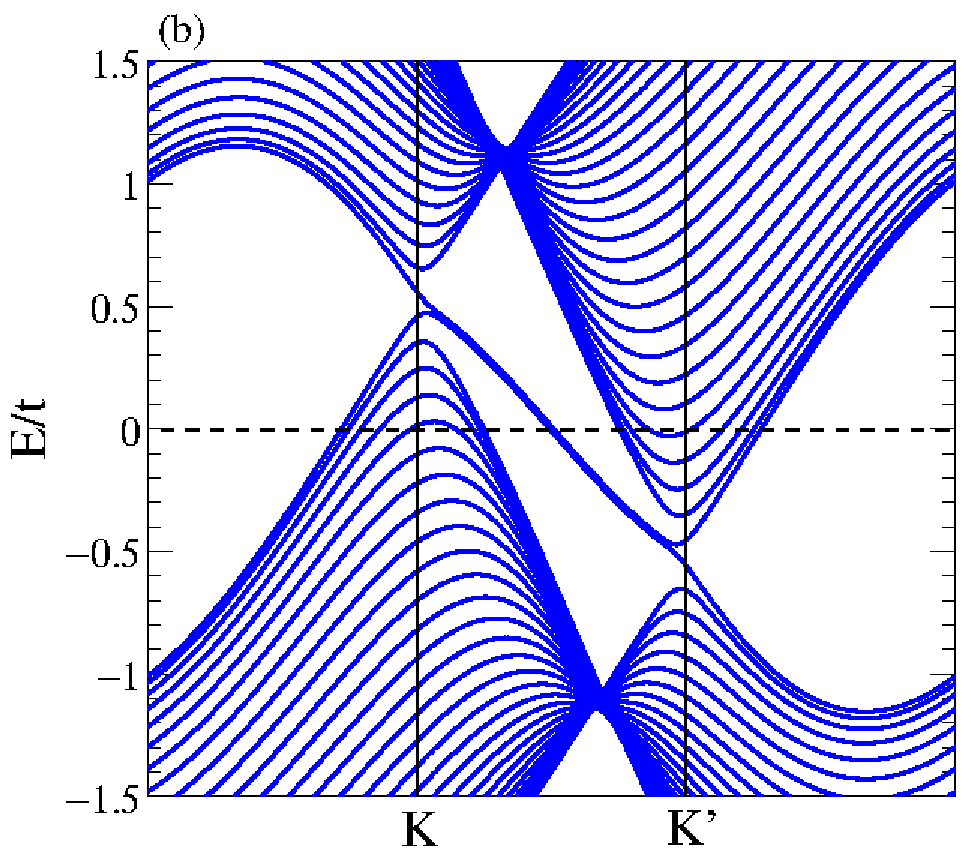}
\includegraphics[width=0.4\columnwidth]{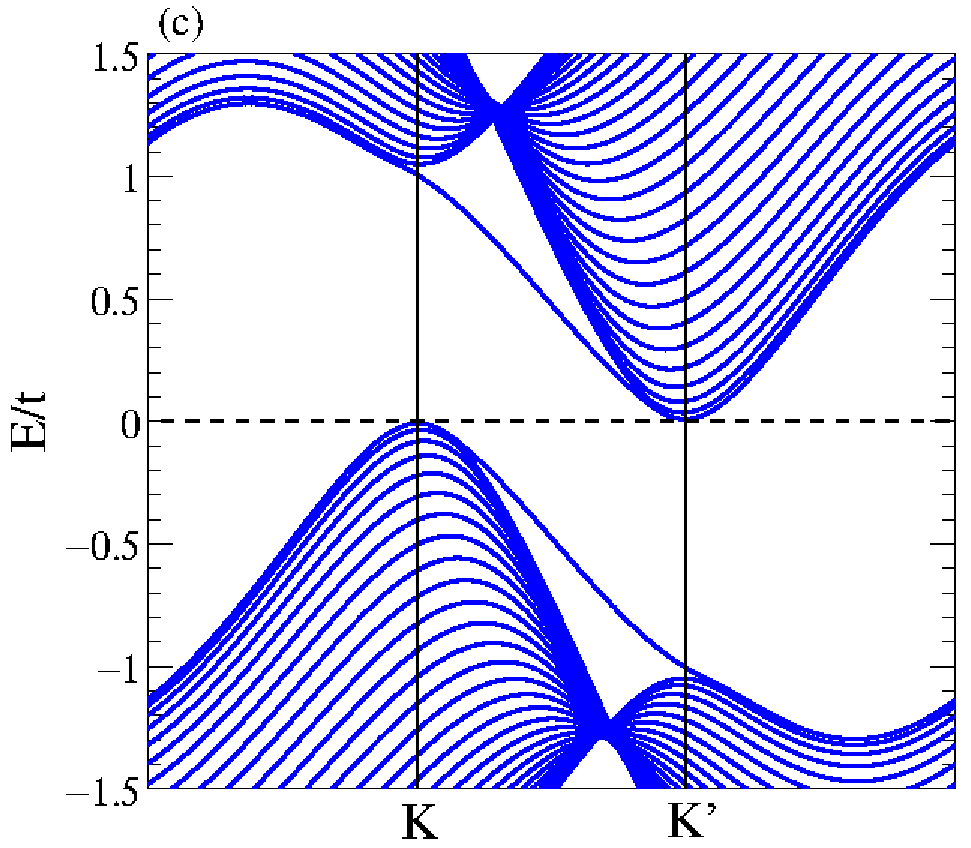}
\includegraphics[width=0.4\columnwidth]{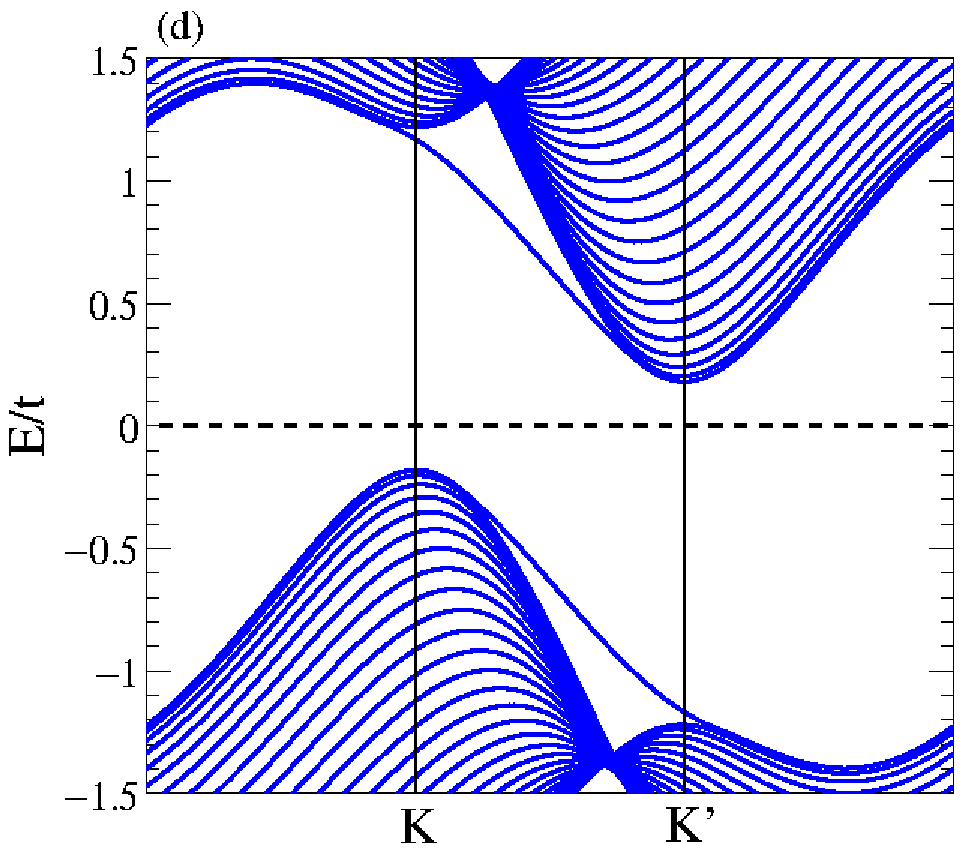}
\end{array}$
\caption{Band structure of the mHM in monolayer graphene nanoribbon with zigzag boundaries and a width $W=60$ atoms. Calculations are done for $t_2=0.1t$, $\Phi=\frac{\pi}2$, (a) $M=0$, (b) $M=0.1t_2$, (c) $M=M_c\equiv 3\sqrt{3}t_2$, and (d) $M_2=4\sqrt{3}t_2$.}
\label{mass-ML}
\centering
$\begin{array}{cccc}
\includegraphics[width=0.4\columnwidth]{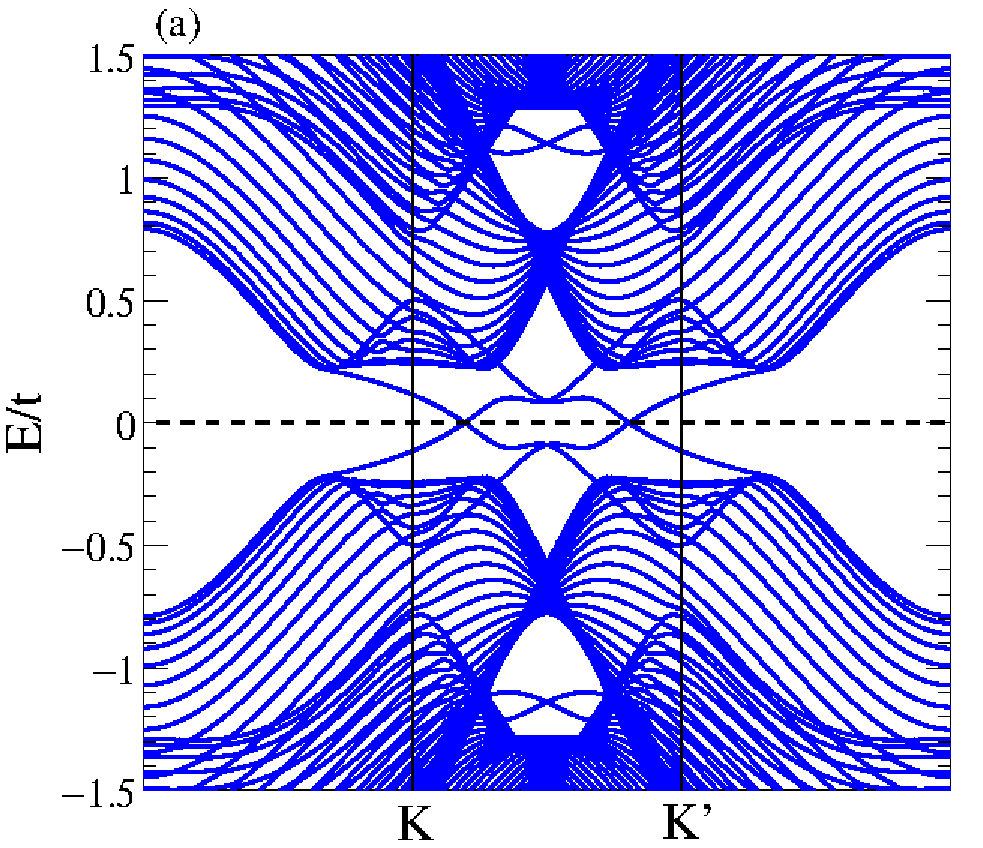}
\includegraphics[width=0.4\columnwidth]{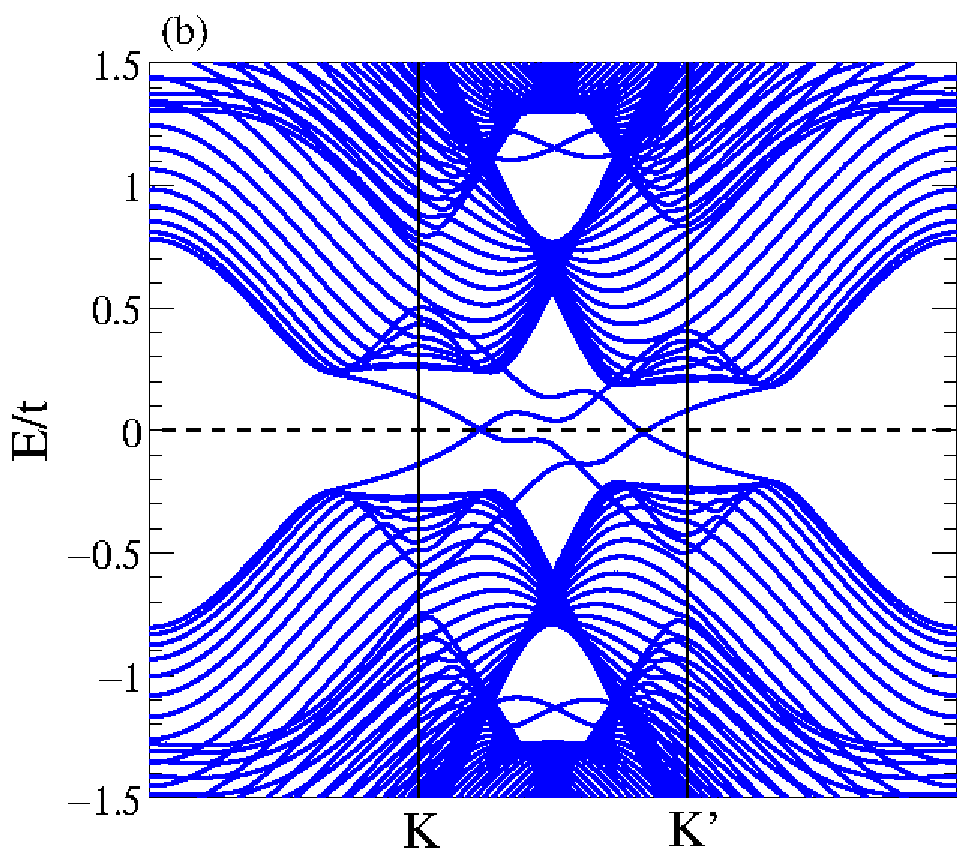}
\includegraphics[width=0.4\columnwidth]{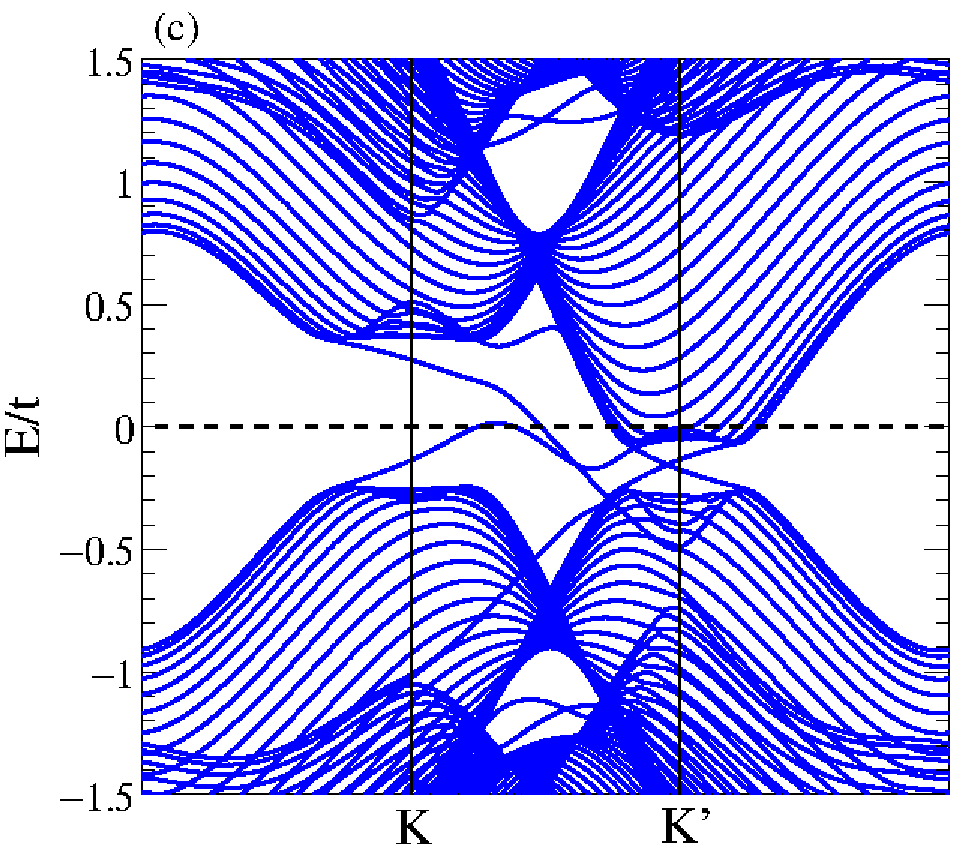}
\includegraphics[width=0.4\columnwidth]{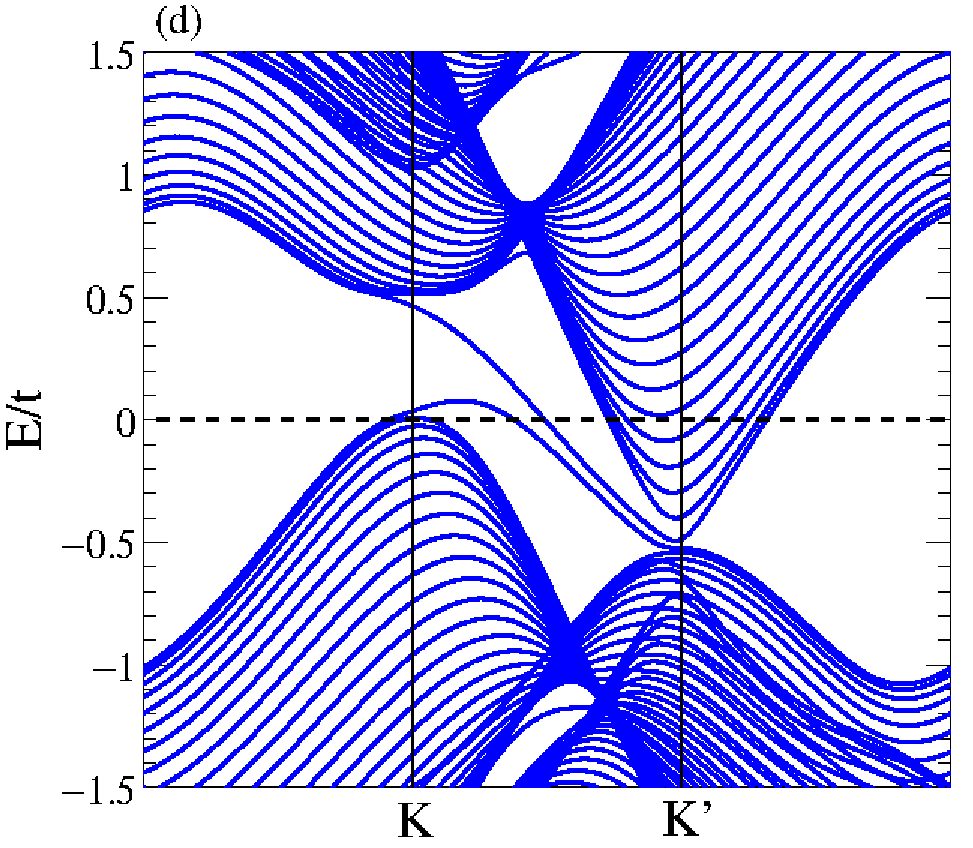}\\
\includegraphics[width=0.4\columnwidth]{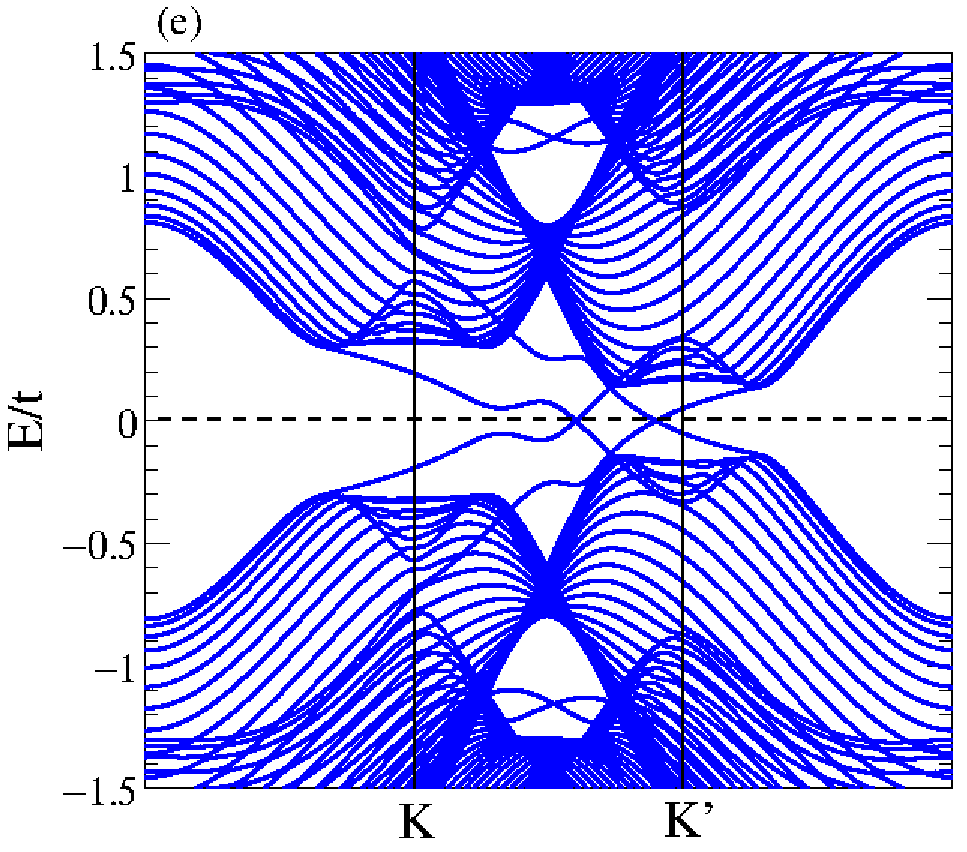}
\includegraphics[width=0.4\columnwidth]{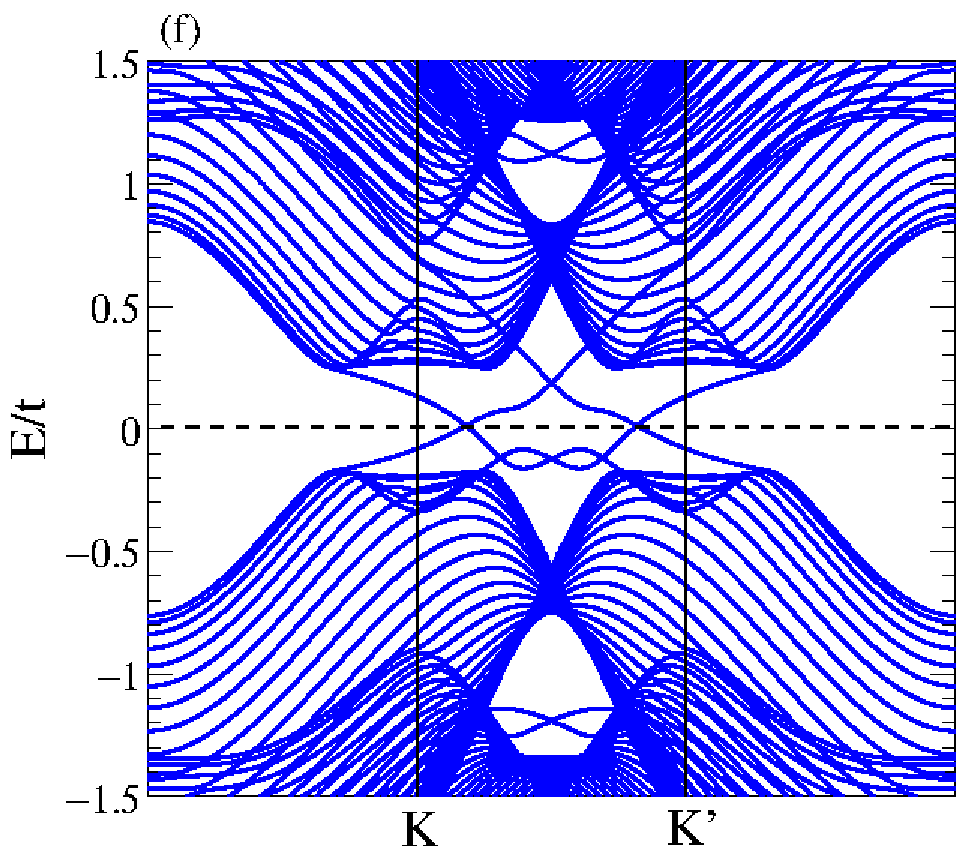}
\includegraphics[width=0.4\columnwidth]{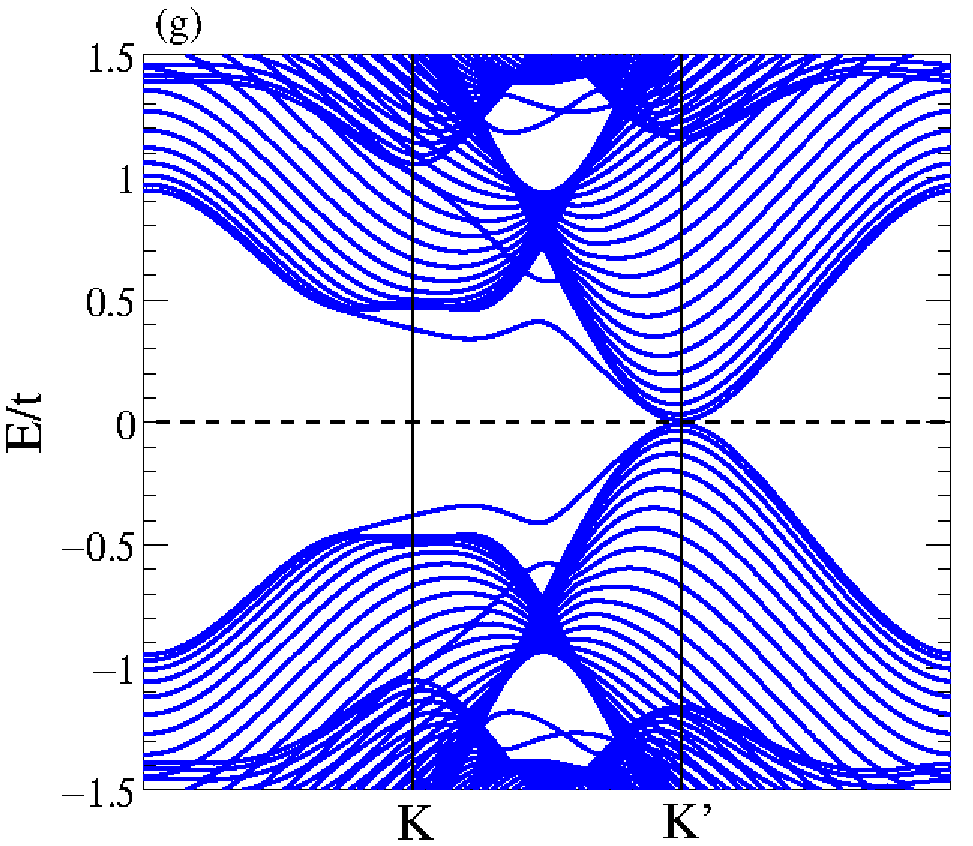}
\includegraphics[width=0.4\columnwidth]{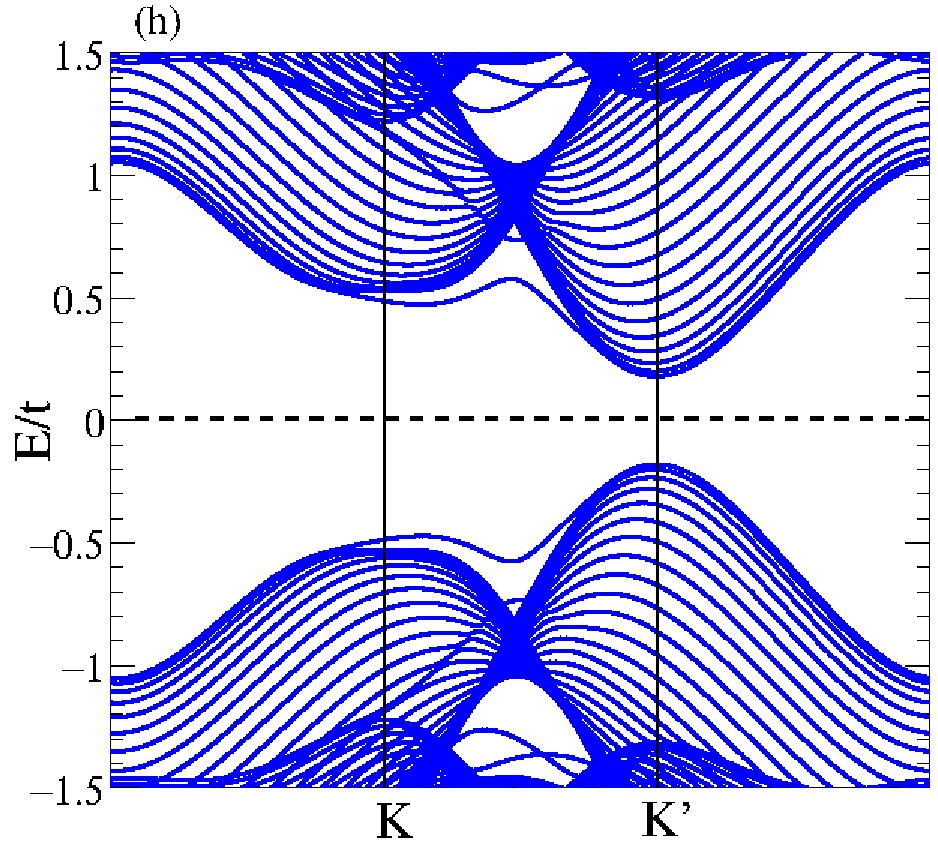}
\end{array}$
\caption{Band structure of the mHM in AB stacked bilayer zigzag ribbons of a width $W=60$ atoms with $t_2=0.1t$, $t_{\perp}=0.5 t$, $\Phi_1=-\Phi_2=\frac{\pi}2$, (a) $M_1=M_2=0$, (b) $M_1=0$, $M_2=t_2$, (c) $M_1=0$, $M_2=3\sqrt{3}t_2$ (d) $M_1=0$, $M_2=6\sqrt{3}t_2$, (e) $M_1=M_2=\sqrt{3}t_2$, (f) $M_1=-M_2=\sqrt{3}t_2$, (g) $M_1=M_2=3\sqrt{3}t_2$, (h) $M_1=M_2=4\sqrt{3}t_2$.}
\label{mass}
\end{figure*}
\end{widetext}

%\pagebreak 


\begin{thebibliography} {200}

\bibitem{Hasan-Rev} M. Z. Hasan and C. L. Kane, Rev. Mod. Phys. {\bf 82}, 3045 (2010).

\bibitem{Qi-Rev} X.-L. Qi and S.-C. Zhang, Rev. Mod. Phys. {\bf 83}, 1057 (2011).

\bibitem{Bansil} A. Bansil, H. Lin, and T. Das, Rev. Mod. Phys. {\bf 88}, 021004 (2016).

\bibitem{AQHE-rev} K. He, Y. Wang and Q.-K. Xue, Annu. Rev. Condens. Matter Phys. {\bf 9}, 329 (2018).

\bibitem{Yang} Z. Yang, F. Gao, X. Shi, X. Lin, Z. Gao, Y. Chong, and B. Zhang, Phys. Rev. Lett. {\bf 114}, 114301 (2015).

\bibitem{Khan} A. Khanikaev, R. Fleury, S. Mousavi, and A. Alu, Nat. Commun. {\bf 6}, 8260 (2015).

\bibitem{Ni} X. Ni, C. He, X.-C. Sun, X.-P. Liu, M.-H. Lu, L. Feng and Y.-F. Chen,   New J. Phys. {\bf 17} 053016 (2015).

\bibitem{Esslinger} G. Jotzu, M. Messer, R. Desbuquois, M. Lebrat, T. Uehlinger, D. Greif and T. Esslinger, Nature {\bf 515}, 237 (2014).

\bibitem{Kee} H.-S. Kim and H-Y Kee, npj Quantum Materials, {\bf 2}, 20 (2017).

\bibitem{acoustic} Y. Ding, Y. Peng, Y. Zhu, X. Fan, J. Yang, B. Liang, X. Zhu, X. Wan, and J. Cheng, Phys. Rev. Lett. {\bf 122}, 014302 (2019).

\bibitem{Serlin1} M. Serlin, C. L. Tschirhart, H. Polshyn, Y. Zhang, J. Zhu, K. Watanabe, T. Taniguchi, L. Balents, and A. F. Young, Science {\bf 367}, 900 (2020).

\bibitem{Serlin2} C. L. Tschirhart, M. Serlin, H. Polshyn, A. Shragai, Z. Xia, J. Zhu, Y. Zhang, K. Watanabe, T. Taniguchi, M. E. Huber, and A. F. Young, Science {\bf 372}, 1323 (2021).

\bibitem{resistance} Y. Okazaki, T. Oe, M., Kawamura, R. Yoshimi, S. Nakamura, S. Takada, M. Mogi, K. S. Takahashi, A. Tsukazaki, M. Kawasaki, Y. Tokura and N.-H. Kaneko {\it et al.}, Nat. Phys. {\bf 18}, 25 (2022).

\bibitem{tuneC}Z. Li, Y. Han, and Z. Qiao, Phys. Rev. Lett. {\bf 129}, 036801 (2022).

\bibitem{Haldane} F. D. M. Haldane, Phys. Rev. Lett. {\bf 61}, 2015 (1988).

\bibitem{Franz} E. Colom\'es and M. Franz, Phys. Rev. Lett. {\bf 120}, 086603 (2018).

\bibitem{Dutta} S. Bhattacharjee, S.Bandyopadhyay, D. Sen, and A. Dutta, Phys. Rev. B {\bf 103}, 224304 (2021).

\bibitem{Kane} C. L. Kane and E. J. Mele, Phys. Rev. Lett. {\bf 95}, 226801 (2005); {\textit ibid}, 146802 (2005).

\bibitem{AZ} A. Altland, and M. R. Zirnbauer, Phys. Rev. B {\bf 55}, 1142 (1997).

\bibitem{mHM symm1} The system breaks TRS and parity $\mathcal{P}$ even for $M=0$ and it is invariant under $\mathcal{TP}$ transformation for $M=0$. For $\Phi\equiv \frac {\pi} 2\mod{[\pi]}$ and $M=0$, $H_{mH}(\mathbf{k})$ the Hamiltonian given by Eq.~\ref{mHMmL} conserves charge-conjugation symmetry as the HM.


\bibitem{HM symm} For $\Phi\equiv \frac {\pi} 2\mod{[\pi]}$ and $M=0$, the HM switches to the D class~\cite{AZ,AZ2} where the charge-conjugation symmetry is conserved. Moreover, in the absence of Semenoff mass, the HM conserves parity $\mathcal{P}=\sigma_x$: $\mathcal{P}^{\dagger} H_{H}(\mathbf{k})\mathcal{P}=H_{H}(-\mathbf{k})$ but is not invariant under $\mathcal{TP}$ transformation.

\bibitem{SOC} T. Frank, P. H\"{o}gl, M. Gmitra, D. Kochan, and J. Fabian, Phys. Rev. Lett. {\bf 120}, 156402 (2018).

\bibitem{AZ2}C.-K. Chiu, J. C. Y. Teo, A. P. Schnyder, and S. Ryu, Rev. Mod. Phys. {\bf 88}, 035005 (2016).

\bibitem{footnote sym} In the case of opposite complex phases $\Phi_1=-\Phi_2$ and masses $M_1=-M_2$, the system is invariant under inversion $\mathcal{P}=\sigma_x\tau_x$. Moreover, in the absence of Semenoff masses and for $\Phi_1=-\Phi_2=\pm\frac{\pi}2$, the energy-shift terms ($a^0_{l,\mathbf{k}}=0$ (Eq.~\ref{al})) vanish, and the system becomes invariant under charge conjugation, which results in a $\mathcal{CP}=\tau_x\sigma_y K$ symmetry.

\bibitem{footnote Eg}In the presence of the Semenoff masses, the energy-spectrum has no more the particle-hole symmetry since the charge-conjugation symmetry is broken. However, it keeps the inversion symmetry for $M_1=-M_2$.

\bibitem{McCann} E. McCann and V. I. Fal'ko, Phys. Rev. Lett. {\bf 96}, 086805 (2006).

\bibitem{footnote}Taking into account the Semenoff masses, a simple analytical expression of the energy spectrum can be derived for $\Phi_1=-\Phi_2=\frac{\pi}2$ and $M_1=M_2=M$, where the charge-conjugation and the parity symmetries are broken. The $A_{\mathbf{k}}$ and $B_{\mathbf{k}}$ terms given by Eq.~(\ref{ABk}) become: $A_{\mathbf{k}}=a^2_{\mathbf{k}}+|f_{\mathbf{k}}|^2+2t_{\perp}^2+M^2$ and $B_{\mathbf{k}}=2\sqrt{|f_{\mathbf{k}}|^2\left(a^2_{\mathbf{k}}+t_{\perp}^2\right)+t_{\perp}^4+a^2_{\mathbf{k}}M^2+2a_{\mathbf{k}}t_{\perp}^2M}$.

\bibitem{Lowdin} P.‐O. L\"owdin, J. Chem. Phys. {\bf 19}, 1396 (1951).

\bibitem{Sticlet}D. Sticlet, F. Piéchon, J.-N. Fuchs, P. Kalugin, and P. Simon, Phys. Rev. B {\bf 85}, 165456 (2012).

\bibitem{SC} N. Batra, S. Nayak, and S. Kumar, Phys. Rev. B {\bf 100}, 214517 (2019).

\bibitem{Private-elec} Y. Chong and Y. Yang, private communication.

\bibitem{Yang21} Y. Yang, D. Zhu, Z. Hang, and Y. Chong, Sci. China Phys. Mech. Astron. {\bf 64}, 257011 (2021).

\bibitem{Private-pho} B. Zhang, private communication

\bibitem{Zhou} P. Zhou, G.-G. Liu, Y. Yang, Y.-H. Hu, S. Ma, H. Xue, Q. Wang, L. Deng, and B. Zhang
Phys. Rev. Lett. {\bf 125}, 263603 (2020).

\bibitem{Mukherjee} R. Mukherjee,H. J. Chuang, M. R. Koehler, N. Combs, A. Patchen, Z. X. Zhou, and D. Mandrus, Phys. Rev. Applied {\bf 7}, 034011 (2017).

\bibitem{Chen} K.Chen, D. Kiriya, M. Hettick, M. Tosun, T.-J. Ha, S. Madhvapathy, S. Desai, A.Sachid, and A. Javey, APL Materials 2, 092504 (2014).

\bibitem{Ying} Z. Ying, S. Zhang, B. Chen, B. Jia, F. Fei, M. Zhang, H. Zhang, X. Wang, and F. Song, Phys. Rev. B {\bf 105}, 085412 (2022).

\bibitem{Zahid} J-X. Yin, S. H. Panand M. Z. Hasan, Nature Review Physics {\bf 3}, 249 (2021).

\bibitem{Choi} Y. Choi, H. Kim, Y. Peng, A. Thomson, C. Lewandowski,
R. Polski, Y. Zhang, H. S. Arora, K. Watanabe, T. Taniguchi, J. Alicea and S. Nadj-Perge1, Nature {\bf 589}, 536 (2021).

\bibitem{Pablo2}A. T. Pierce, Y. Xie, J. M. Park, E. Khalaf, S. H. Lee, Y. Cao, D. E. Parker, P. R. Forrester, S. Chen, K. Watanabe, T. Taniguchi, A. Vishwanath, P. Jarillo-Herrero, and A. Yacoby, Nature Physics, {\bf 17}, 1210 (2021).


\bibitem{Kim} S. Kim, J. Schwenk, D. Walkup, Y. Zeng, F. Ghahari,
S. T. Le, M. R. Slot, J. Berwanger, S. R. Blankenship, K. Watanabe,
T. Taniguchi, F. J. Giessibl, N. B. Zhitenev, C. R. Dean, and J. A. Stroscio, Nature Communications {\bf 12}, 2852 (2021).

\bibitem{Claudia} N. B. M. Schr\"{o}ter, S. Stolz, K. Manna, F. de Juan, M. G. Vergniory, J. A. Krieger, D. Pei, T. Schmitt, P. Dudin, T. K. Kim, C. Cacho, B.  Bradlyn, H. Borrmann, M. Schmidt, R. Widmer, V. N. Strocov, C. Felser, Science {\bf 369}, 179 (2020).

\bibitem{bearded} Stacking-induced Chern insulator is expected to occur in mHM on AB bilayer ribbons with bearded edges regarding the presence of zero energy states as in the zigzag ribbons. However, we did not focus on this type of ribbons since they are
less stable compared to the zigzag and the armchair ones~\cite{bearded-Ref}.

\bibitem{bearded-Ref} M. Kohmoto and Y. Hasegawa, Phys. Rev. B {\bf 76}, 205402 (2007).

\bibitem{Ctunue} Y.-F. Zhao, R. Zhang, R. Mei, L. -J. Zhou, H. Yi, Y.-Q. Zhang, J. Yu, R. Xiao, K. Wang, N. Samarth, M. H. W. Chan, C.-X. Liu, and  C.-Z. Chang, Nature {\bf 588},  419 (2020).

\bibitem{Guy} M. Manna\"{i}, G.  Trambly de laissardi\`ere, S. Haddad, J.-N. Fuchs and F. Piéchon, in preparation. 

\bibitem{Herrero} Y. Cao, V. Fatemi, S. Fang, K. Watanabe, T. Taniguchi, E. Kaxiras, and P. Jarillo-Herrero, Nature, {\bf 556}, 43 (2018).

\end{thebibliography}
\end{document}